# Optimized CFD modelling and validation of radiation section of an industrial top-fired steam methane reforming furnace

Mustafa Tutar[a], Cihat Emre Üstün[b,*], Jose Miguel Campillo-Robles[c], Raquel Fuente[b], Silvia Cibrián[d], Ignacio Arzua[d], Arturo Fernández[e], Gabriel A. López[c]

[a] *Department of Energy Engineering, Ankara University, 06830 Ankara, Turkey*
[b] *Applied Mathematics, Faculty of Engineering, University of the Basque Country UPV/EHU, Plaza Ingeniero Torres Quevedo, 1, Bilbao 48013, Spain*
[c] *Fisika Saila, Zientzia eta Teknologia Fakultatea, UPV/EHU,*
[d] *Petróleos del Norte, San Martín 5, 48550 Muskiz, Spain*
[e] *Petronor Innovación, San Martín 5, 48550 Muskiz, Spain*

ABSTRACT

The present study proposes an optimized computational fluid dynamics (CFD) modelling framework to provide a complete and accurate representation of combustion and heat transfer phenomena in the radiation section of an industrial top-fired steam methane reforming (SMR) furnace containing 64 reforming tubes, 30 burners and 3 flue-gas tunnels. The framework combines fully-coupled appropriate furnace-side models with a 1-D reforming process-side model. Experimental measurements are conducted in terms of outlet temperatures at the flue-gas tunnels, point-wise temperature distributions at the panel walls, and inside the reforming tube collectors which are placed at the refinery plant of Petronor. The final results are compared with the experimental data for validation purpose. The proposed fully coupled 3-D CFD modeling framework, which utilizes a detailed chemical-kinetic combustion mechanism, reproduces well basic flow features including pre-mixed combustion process, downward movement of flue-gas in association with large recirculation zones, radiative heat transfer to the reforming tubes, composition profiles along the reaction core of the reforming tubes, temperature non-uniformities, and fluctuating characteristics of heat flux. The reported non-uniform heat and temperature distributions might be optimized by means of the operating parameters in order to avoid a negative impact on furnace balancing and performance.

## 1. Introduction

Hydrogen demand is increasingly rising worldwide (around 5–10 % annually), and it is expected to continue growing in the next decades (Tran et al., 2018). Indeed, hydrogen is considered a promising energy carrier, which could replace fossil fuels in most of their current applications (Töpler et al., 2016). Hydrogen has several advantages (Stolten et al., 2016): clean combustion, high gravimetric energy density and wide application in the fields of petroleum, chemical, electronics, metallurgy, etc. Nowadays, hydrogen is principally used in petroleum refineries, chemical manufacturing and in industrial and energy markets. Actually, hydrogen is one of the most important raw materials for petroleum industries for hydrotreating and hydrocracking (Rostrup-Nielsen and Christiansen, 2011). For example, hydrogen is used to convert crude oil into a variety of high valuable products, like gasoline, jet fuel and diesel, but it is also important in the production of ammonia, methanol and synfuels. Moreover, it seems that the development of a new hydrogen economy paradigm is mainly associated with the use of fuel cells (Abdi et al., 2017).

Although there are other ways to obtain hydrogen, today, it is produced mainly by two processes (Lipman et al., 2019): reforming and electrolysis (Godula-Jopek, 2015). Nowadays, reforming processes of fossil fuels are the dominant processes for hydrogen production. Typical feedstock is natural gas (methane), but liquid hydrocarbons such as gasoline or methanol can also be used. Steam methane reforming (SMR) is the most dominant method worldwide for hydrogen production (Abbas, 2014), because it is of relative low cost compared with other available technologies (Sankir et al., 2017). Indeed, SMR hydrogen production

* Corresponding author.
*E-mail address:* custun001@ikasle.ehu.eus (C.E. Üstün).

ranges from 90 to 95 % of the total world hydrogen production (Pashchenko, 2019).

SMR was introduced into the industry in the 1930s. Therefore, it is a well-established and mature technology (Murkin and Brightling, 2016). The SMR furnace has two main modules: a reactor and a combustor. In the reactor, methane gas and superheated steam are converted into syngas (a mixture of gases among those, which are hydrogen, carbon monoxide and carbon dioxide) in the presence of a nickel-based catalyst, by a sequence of net endothermic reactions at high temperature, above 753 K (Shah et al., 2017). These reactions occur in catalyst-filled tubes placed inside the combustor or furnace chamber, where combustion reactions of natural gas with inlet air take place to obtain the required heat-release. In order to obtain a high methane conversion and reduce the coke deposition on the catalyst, superheated steam is supplied to the reformer more than the stoichiometric amount. The steam-to-methane ratio is an essential parameter that must be optimized according to working temperature and pressure values (Yin et al., 2006). While the recommended values of steam-to-methane ratio are usually between 2 and 3 for modern furnaces (Pashchenko, 2019), the reported values are in the range of 1.7 to 9 depending on the operational considerations of SMR furnaces (Zecevic and Bolf, 2020, Acuña et al., 1999). The conversion performance from methane to hydrogen usually amounts to 74–85% (Ibrahim, 2018).

There are many mathematical models in the academic and commercial literature to give a detailed description of SMR furnaces. These models differ in their intended use and in the simplifying assumptions they use to describe furnace behavior. It is necessary to have a complete SMR model to analyze the system as a whole and validate both furnace side and process side numerical results with the experimental results. In the last decades, computational fluid dynamics (CFD) modeling of multiphysics systems has observed many important developments due to the progress in numerical methods and the increment of computational capacity (Zhang et al., 2015, Peksen, 2018). CFD has become a cornerstone in the research and development phase of combustion systems. Therefore, CFD is a reliable and robust tool to simulate processes and systems including complex flow phenomena such as combustion and reactive flows (Jurtz et al., 2019, Molarski et al., 2020, Singh et al., 2013). SMR phenomena can be modeled using CFD solvers. Principally, CFD models of SMR furnaces can be divided in three main types (Latham, 2008):

-*Process side models:* These models simulate the kinetics of the reforming catalytic process, and are very useful for industrial unit simulation, optimization, operational monitoring, sensitivity analysis and design. They take into account detailed information about the inlet fluxes, the catalyst and the chemical reaction kinetics (Dixon et al., 2006, Dixon and Partopour, 2020, Inbamrung et al., 2018, Li et al., 2019, Kuncharam and Dixon, 2020). Although process side models can achieve detailed analyses as mentioned, they often use over-simplifying boundary conditions such as applying constant heat flux boundary conditions for the reformer walls unless there is reliable experimental temperature or heat flux profile data (Dixon, 2017, Pashchenko, 2018).

-*Furnace side models*: These models take into account principally combustion and radiation heat transfer. They usually describe the flow patterns and re-circulations inside the furnace (Farnell and Cotton, 2000). One important drawback of these models is not to take actual endothermic reactions happening inside the reforming tubes into consideration which in return may result in incomplete and misleading results. In fact, both radiation and convection mechanisms of heat transfer affect the flue gas flow directions and magnitudes by creating temperature gradients in the flow domain.

-*Complete steam-methane reforming models:* These models, as opposed to previous models, combine process side and furnace side models to capture SMR phenomena in a complete manner. Therefore, when validated against experimental data, these models can be used as a reliable technique in the modelling of industrial SMR furnaces. One of the early attempts of complete SMR modelling studies is achieved by running process side and furnace side models separately and coupling the runs by exchanging data between the models through the boundary conditions (Nielsen and Christiansen, 2002). Inconvenience of this strategy together with unrevealed modelling parameters such as combustion mechanism, mesh and model independency tests does not provide a good insight on the study. Coupling process and furnace side models without the need of running them separately is achieved by Zheng et al., 2010a), Zheng et al., 2010b). However, this work also lacks mesh and model independency studies and assumes the symmetry boundary condition as a sufficient modelling strategy. In recent years, the research group led by Christofides has made remarkable contributions to the literature of modelling of SMR furnaces (Tran et al., 2017b, Lao et al., 2016, Tran et al., 2017a, Tran et al., 2018). Tran et al., 2017b investigate CFD modelling strategies by comparative analyses and create a base CFD model for an industrial hydrogen reformer. Although comprehensive, the model is computationally expensive due to 3-D process side modelling technique. Moreover, industrial-scale CFD models can be utilised as a base to develop novel control schemes for furnace operation (Lao et al., 2016). When validated, CFD models can also be used to optimize the furnace operation by means of temperature balancing and furnace-side feed distribution (Tran et al., 2017a, Tran et al., 2018).

Motivated by the above model considerations, the present study utilizes a finite volume method (FVM) based CFD modelling approach that combines fully-coupled appropriate furnace side models with a 1-D process side model to provide a complete and accurate representation of basic flow/heat transfer features in the firebox of an industrial SMR furnace. With this respect, some turbulence-combustion-radiation models are evaluated to relate the premixed combustion process with the radiation process though a specialized reaction model. The objective is to accurately represent the velocity vector field, heat transfer and temperature distributions in the firebox and heat flux/temperature distributions and composition profiles of reacting species in the reforming tubes with reasonably low computational time and cost. For doing this, three different Reynolds averaged Navier-Stokes (RANS) equation-based turbulence models are coupled with either a standard Eddy Break-up (EBU) model or Flamelet Generated Manifold (FGM) model with a sub-model called Turbulent Flame Speed Closure (TFC), Discrete-Ordinates (DO) model with weighted-sum-of-gray-gases (WSGGM) radiation sub-model and a 1-D plug flow Reacting Channel (RC) model. Once the best suitable coupled modelling approach is chosen in comparison with the experimentally measured data, the present study is extended to further thorough investigation of flow field and temperature distributions at different planes of the firebox and the reforming tubes.

## 2. Experimental set-up

The analyzed system is a top-fired SMR furnace placed at the refinery plant of Petronor in Muskiz (Basque Country), which has been operational approximately over 40 years. The geometry of the furnace is shown in Fig. 1.

The full three-dimensional (3-D) furnace which has a length of 9.820 m, width of 5.476 m and height of 11.911 m, consists of a total of 64 reforming tubes in two rows, 30 burners in three rows at the ceiling of the furnace and three flue-gas tunnels for evac-uation. Table 1 shows the technical details of analyzed reforming furnace.

**Table 1**
Technical details of analyzed reforming furnace.

| Technical detail | Value |
|---|---|
| Designed thermal efficiency (%): | 90.2 |
| Power (MW) | 17 |
| Outer diameter of reforming tubes (mm): | 130.7 |
| Inner diameter of reforming tubes (mm): | 99.7 |
| Total length of reforming tubes (m): | 13.41 |
| Heated length of reforming tubes (m): | 11.07 |
| Distance between rows of reforming tubes (m): | 1.83 |
| Centre to center distance of reforming tubes (cm): | 26.67 |
| Emissivity of reforming tubes: | 0.85 (Latham, 2008, Tran et al., 2018) |
| Distance between rows of burners (m): | 1.96 |
| Burner center to tube center distance (m): | 0.915 (inner burner row) 1.040 (outer burner rows) |
| Excess air in the combustion chamber (vol %): | 4.12 |
| Emissivity of furnace inner walls and tunnels: | 0.6 (Tran et al., 2018) |
| Section of evacuation tunnels ($m^2$): | 0.61 × 1.24 |
| Composition of catalyst: | Ni on a calcium aluminate support |

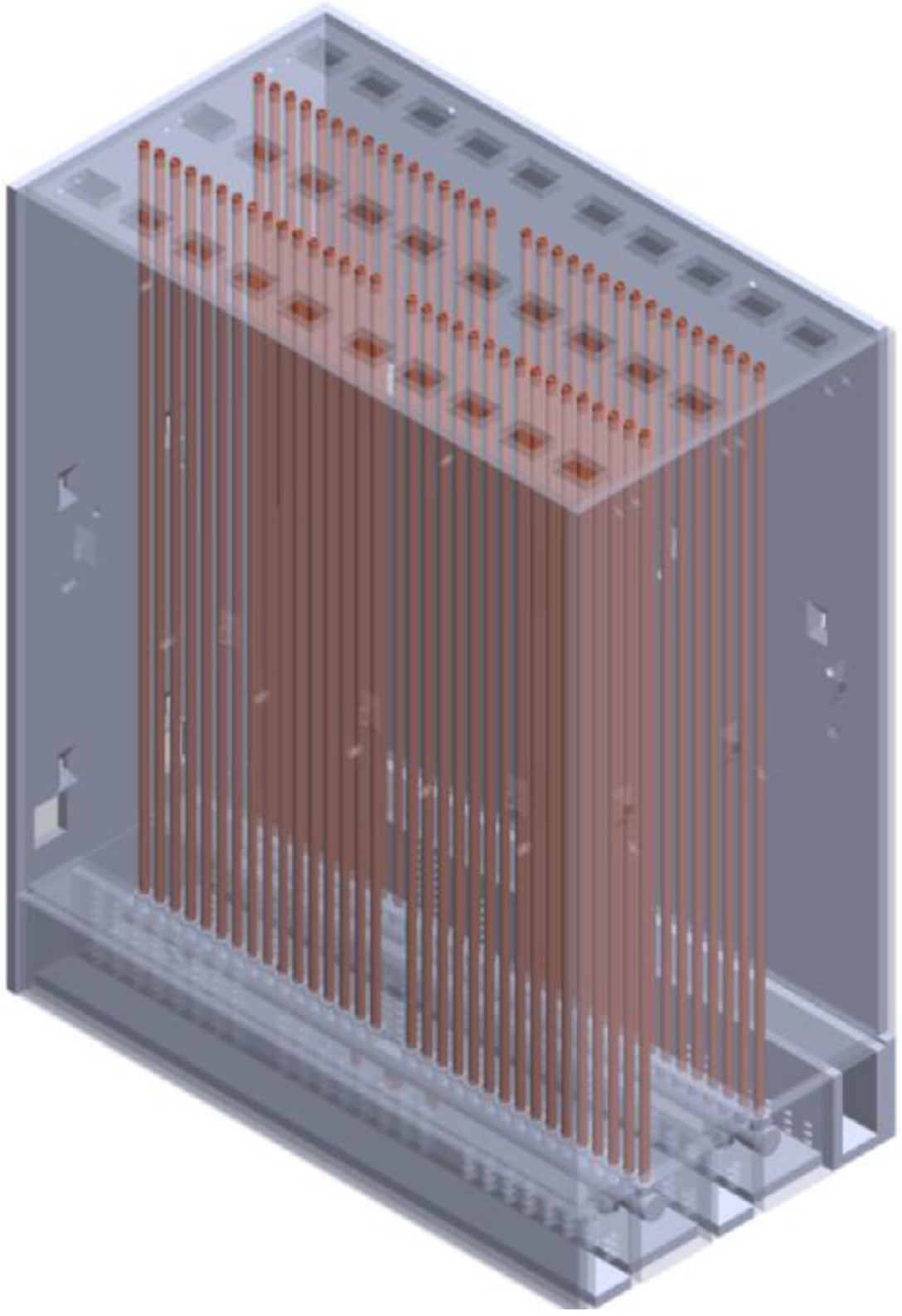

**Fig. 1.** Isometric view of an industrial, top-fired, co-current reformer with 64 reforming tubes, 30 burners and 3 flue-gas tunnels.

Fuel gas is the usual combustible of this furnace, and its average composition appears in Table 2. Fuel gas combustion process has an efficiency of 100 %, due to the excess of oxygen (4.12 %) introduced in the combustion chamber to avoid explosions. Table 3 shows the average composition of the reforming fuel gas, which is mixed with superheated steam to perform the catalysis process inside the reforming tubes.

The reforming fuel gas average composition used in the process side is provided in Table 3. For natural gas combustion, a detailed chemical-kinetic combustion mechanism so-called GRI Mech 3.0 is used (GRI Mech 3.0). In this approach, the number of species and reactions are kept to the minimum needed to describe the systems and the phenomena addressed, thereby minimizing as much as possible the uncertainties in the rate parameters employed.

SMR process, which takes place inside the reforming tubes, can be described with detailed chemical mechanisms in micro scale. A global kinetic mechanism of the SMR which is successfully applied by Tran et al. is utilized in this study (Tran et al., 2017b, Xu and Froment, 1989). The SMR process used in industry is mainly an endothermic reaction, which takes place at a high temperature and high pressure (20–35 bar). SMR reactions under standard conditions can be represented as below:

$$CH_4(g) + H_2O(g) \Leftrightarrow CO\,(g) + 3H_2\ (g)\cdots\ \Delta H^o_{298} = 206\,kJ/mol$$

$$CO(g) + H_2O\ (g) \Leftrightarrow\ CO_2(g) + H_2(g)\cdots\ \Delta H^o_{298} = -41.1\,kJ/mol$$

$$CH_4(g) + 2H_2O(g) \Leftrightarrow\ CO_2(g) + 4H_2\ (g)\cdots\ \Delta H^o_{298} = 164.9\,kJ/mol$$

It can be noticed from the standard enthalpies of the reactions that the first and the third reactions are endothermic while the second reaction is exothermic. The overall reaction is highly endothermic and requires external heat.

There are different sensors positioned in the reforming furnace to measure: temperature, pressure, flow rate and composition. These sensors are used to monitor and control the reforming and combustion processes inside the analyzed furnace. Fig. 2 shows the positions of the temperature sensors used for validation purposes (see also Table 5). Temperature sensors, TS-1 to TS-6 are mounted on the walls of the furnace. Sensors TS-7 to TS-10 are placed inside the reforming tube (RT) outlet collectors. TS-11 and TS-12 are placed at the south and middle outlet boundaries of the firebox, respectively.

Tables 4 and 5 show all of the technical information about the sensors. Measurements of these sensors are used as input for the model and to validate the results of the model. Measurements are performed in a time period of 80 days under typical working conditions of the furnace. Averaged values and their standard deviations of measured parameters are calculated (see Tables 4 and 5).

## 3. Modelling analysis of flow through the full 3-D furnace

### 3.1. Computational domain set-up

A full-scale, 3-D model geometry of the furnace is created and CFD simulations are performed in the corresponding computational flow domain. The idea here is that the entire firebox, i.e. the radiative heat transfer section of the furnace, is modeled in a 3-D manner with a coupled CFD modeling approach to resolve the flow/heat transfer features in a full, 3-D computational flow domain. Top and front views of the furnace are illustrated in Fig. 3 (a) and (b). Heat transfer and flow parameters will be plotted using some lateral planes and frontal planes, which are clearly illustrated in Fig. 3 (a). The heights of the horizontal planes are 9.92 m, 5.92 m and 1.92 m for HP1, HP2 and HP3, respectively and are shown in Fig. 3 (b).

**Table 2**
Average composition of combustion fuel gas.

| Fuel gas composition | Volumetric proportions (%) | Standard deviation (%) | Volumetric flow rate ($m^3\ h^{-1}$) | Mass flow rate ($kg\ h^{-1}$) |
|---|---|---|---|---|
| Methane ($CH_4$) | 43.35 | 9.54 | 461.90 | 429.51 |
| Ethylene ($C_2H_4$) | 0.02 | 0.06 | 0.21 | 0.35 |
| Ethane ($C_2H_6$) | 13.75 | 3.53 | 146.65 | 255.65 |
| Propylene ($C_3H_6$) | 0.14 | 0.16 | 1.49 | 3.64 |
| Propane ($C_3H_8$) | 9.46 | 2.64 | 100.96 | 258.12 |
| Hexane ($C_6H_{14}$) | 0.21 | 0.09 | 2.24 | 11.17 |
| Carbon dioxide ($CO_2$) | 0.08 | 0.04 | 0.85 | 2.17 |
| Nitrogen ($N_2$) | 0.30 | 0.04 | 3.20 | 5.19 |
| Hydrogen ($H_2$) | 28.66 | 7.59 | 305.45 | 35.77 |
| Isobutane ($i$-$C_4H_{10}$) | 1.40 | 0.47 | 14.91 | 50.24 |
| Normal Butane ($n$-$C_4H_{10}$) | 2.29 | 0.79 | 24.39 | 82.18 |
| Isopentene ($i$-$C_5H_{10}$) | 0.34 | 0.12 | 3.62 | 15.15 |
| TOTAL | 100.00 | NA | 1065.88 | 1149.13 |

**Table 3**
Average composition of reforming fuel gas.

| Fuel gas composition | Volumetric proportions (%) | Standard deviation (%) | Volumetric flow rate ($m^3\ h^{-1}$) | Mass flow rate ($kg\ h^{-1}$) |
|---|---|---|---|---|
| Methane ($CH_4$) | 91.70 | 1.02 | 2584.67 | 1850.32 |
| Ethylene ($C_2H_4$) | 0.01 | 0.04 | 0.28 | 0.35 |
| Ethane ($C_2H_6$) | 2.78 | 1.15 | 79.03 | 106.06 |
| Propane ($C_3H_8$) | 0.40 | 0.08 | 11.25 | 22.14 |
| Hexane ($C_6H_{14}$) | 0.06 | 0.08 | 1.69 | 6.49 |
| Carbon dioxide ($CO_2$) | 0.15 | 0.07 | 4.22 | 8.29 |
| Nitrogen ($N_2$) | 0.10 | 0.08 | 2.81 | 3.52 |
| Hydrogen ($H_2$) | 4.64 | 0.89 | 131.06 | 11.82 |
| Isobutane ($i$-$C_4H_{10}$) | 0.08 | 0.04 | 2.25 | 5.84 |
| Normal Butane ($n$-$C_4H_{10}$) | 0.08 | 0.04 | 2.25 | 5.84 |
| TOTAL | 100.00 | NA | 2819.51 | 2020.66 |

**Table 4**
Experimental measurements of the sensors used as input parameters to the CFD model.

| Magnitude | Sensor Type | Location | | Average Measurement | Standard Deviation (%) | Average Value |
|---|---|---|---|---|---|---|
| Temperature | Type K thermocouple TTMUX MTL-831/838 | Inlet of the two rows of reforming tubes | North row | 676 K | 1.63 | 676.4 K |
| | | | South row | 676.8 K | 1.63 | |
| Gauge Pressure | Emerson 3051CD3 | Inlet of the three rows of fuel gas -injection to the burners | North row | 71.59 kPa | 18.15 | 57.53 kPa |
| | | | Central row | 48.05 kPa | 33.29 | |
| | | | South row | 52.96 kPa | 22.65 | |
| Gauge Pressure | Emerson 3051CG5 | Inlet of the reforming tubes | | 1872.1 kPa | 4.33 | NA |
| Volumetric flow rate | Orifice plate and Emerson 3051CD2 | Inlet of fuel gas to the burners | | 1520.2 $Nm^3\ h^{-1}$ | 8.48 | NA |
| Volumetric flow rate | Orifice plate and Emerson 3051CD2 | Inlet of gas to reforming tubes | | 2812.5 $Nm^3\ h^{-1}$ | 17.23 | NA |
| Mass flow rate | Orifice plate and Emerson 3051CD2 | Inlet of steam to reforming tubes | | 14177.0 kg $h^{-1}$ | 8.97 | NA |
| Composition exhausted gases – Oxygen proportion | Emerson WC3000 (zirconium oxide) | End of evacuation tunnels | Central Tunnel | 4.47 % | 1.40 | 4.12 % |
| | | | North Tunnel | 3.76 % | 1.40 | |

In the present study, realistic boundary conditions according to furnace working conditions are imposed at the defined boundaries of the full model. The only assumption for the simulations here is related to the imposed heat flux value due to the heat transfer behavior of the outer walls of the furnace. Due to the fact that weather conditions and working conditions of the furnace are not stable, it is not possible to define a “conjugate heat transfer” thermal boundary condition on the refractory walls which may account for heat losses of the furnace. However, it is stated several times in the literature that a “constant heat flux” boundary condition can be applied to account for a heat loss of 2 percent of produced thermal energy by the burners (Latham, 2008, Tran et al., 2017b,

**Table 5**
Temperature sensors (Type K thermocouple TTMUX MTL-831/838) and measurements used for validation of the CFD model.

| Sensor Code | Location | | Average Measurement (K) | Standard Deviation (%) | Average Value (K) |
|---|---|---|---|---|---|
| TS-1 | Combustion chamber - 0.35 m below the burners | North East | 1363.6 | 3.02 | 1347.0 |
| TS-2 | | North West | 1339.2 | 3.24 | |
| TS-3 | | South East | 1354.1 | 2.64 | |
| TS-4 | | South West | 1331.1 | 3.02 | |
| TS-5 | Combustion chamber - 3.35 m below the burners | North East | 1123.5 | 3.33 | 1118.8 |
| TS-6 | | North West | 1114.1 | 3.12 | |
| TS-7 | Outlet manifolds of reforming tubes | North East | 1027.2 | 1.26 | 1016.5 |
| TS-8 | | North West | 1003.7 | 1.09 | |
| TS-9 | | South East | 1024.8 | 1.17 | |
| TS-10 | | South West | 1010.3 | 1.25 | |
| TS-11 | End of evacuation tunnels | Central | 1062.7 | 1.40 | 1063.9 |
| TS-12 | | South | 1065.0 | 1.48 | |

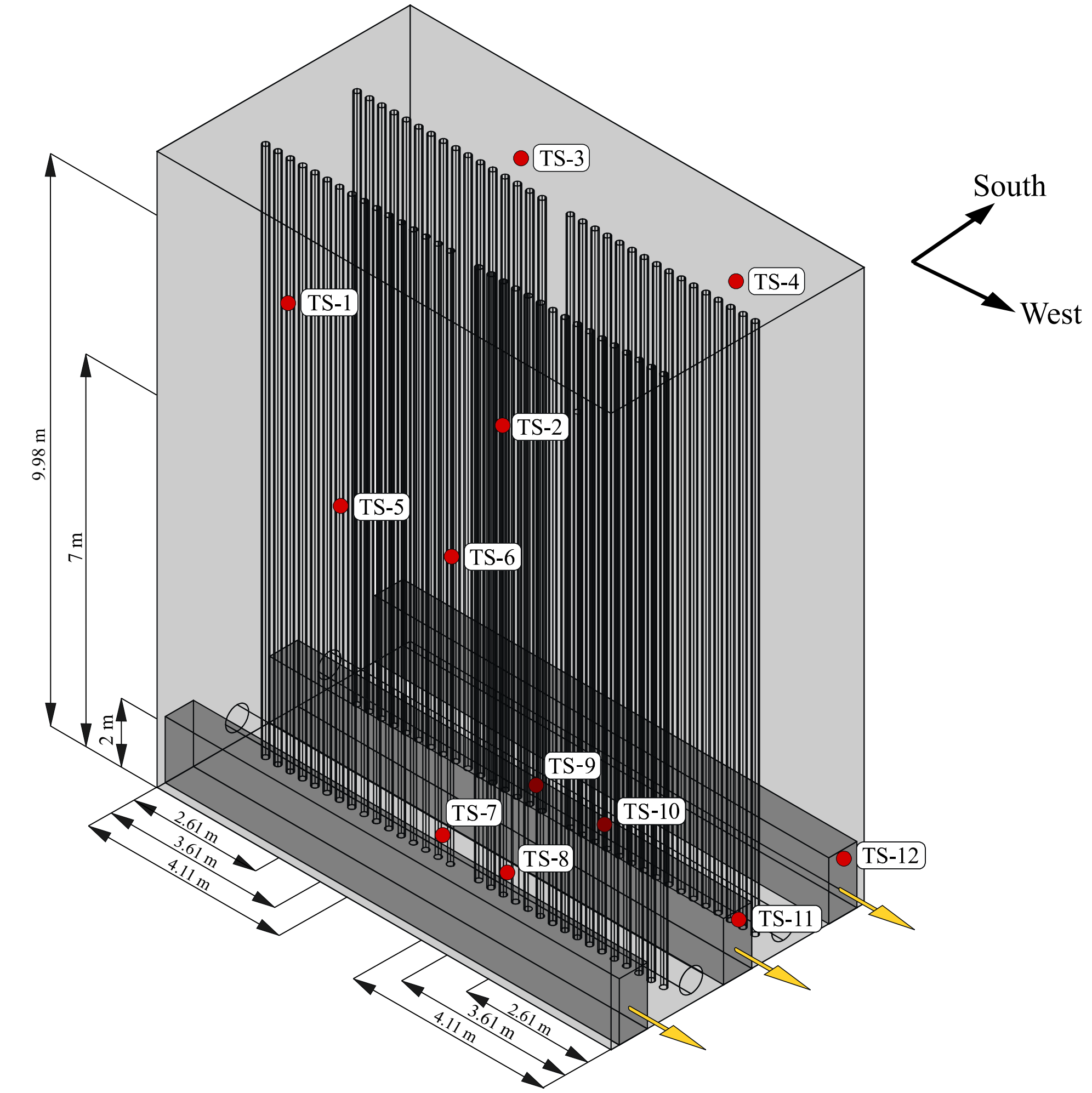

**Fig. 2.** Positions of the temperature sensors in the furnace.

Darvishi and Zareie-Kordshouli, 2017). Further information about the boundary conditions is given in Section 3.3.2.

### 3.2. Mesh set-up

The computational domain illustrated in Fig. 4 is divided into small and discrete sub-domains also known as grids (a collection of grids is referred to as a mesh), within which spatial variations are, though not negligible, significantly less drastic than those in the overall domain. Then, the reformer mathematical model is discretized and numerically solved within each grid to characterize the fluid-flow and temperature fields. The three-dimensional (3-D) computational domain is constructed of non-uniformly spaced unstructured mesh system up to 16,800,000 cells (polyhedral + prism layer), with very fine mesh resolution near the walls surfaces, i.e. the surfaces of the furnace, tubes, and burners as seen from 3-D mesh layout of the flow domain in Fig. 4 (a) – (d). A polyhedral meshing algorithm is used for mesh generation. Polyhedral meshes are relatively easy and efficient to build, requiring no more surface preparation than the equivalent tetrahedral mesh. They also contain approximately five times fewer cells than a tetrahedral mesh for a given starting surface. The mesh near the solid boundaries is significantly refined with the use of boundary meshes (five prism layers are adopted) by using two-layer all $y^+$ wall treat-ment approach which uses blended wall functions and provides valid boundary conditions for flow, energy and turbulence quantities for a wide range of near-wall mesh densities to accurately simulate the large velocity and temperature gradients in the very close proximity. The first grid point adjacent to the tube wall surface is placed at around $y^+$ 0.2 – 5.0 for a selected final mesh system of 16,800,000 cells, as seen in Fig. 5.

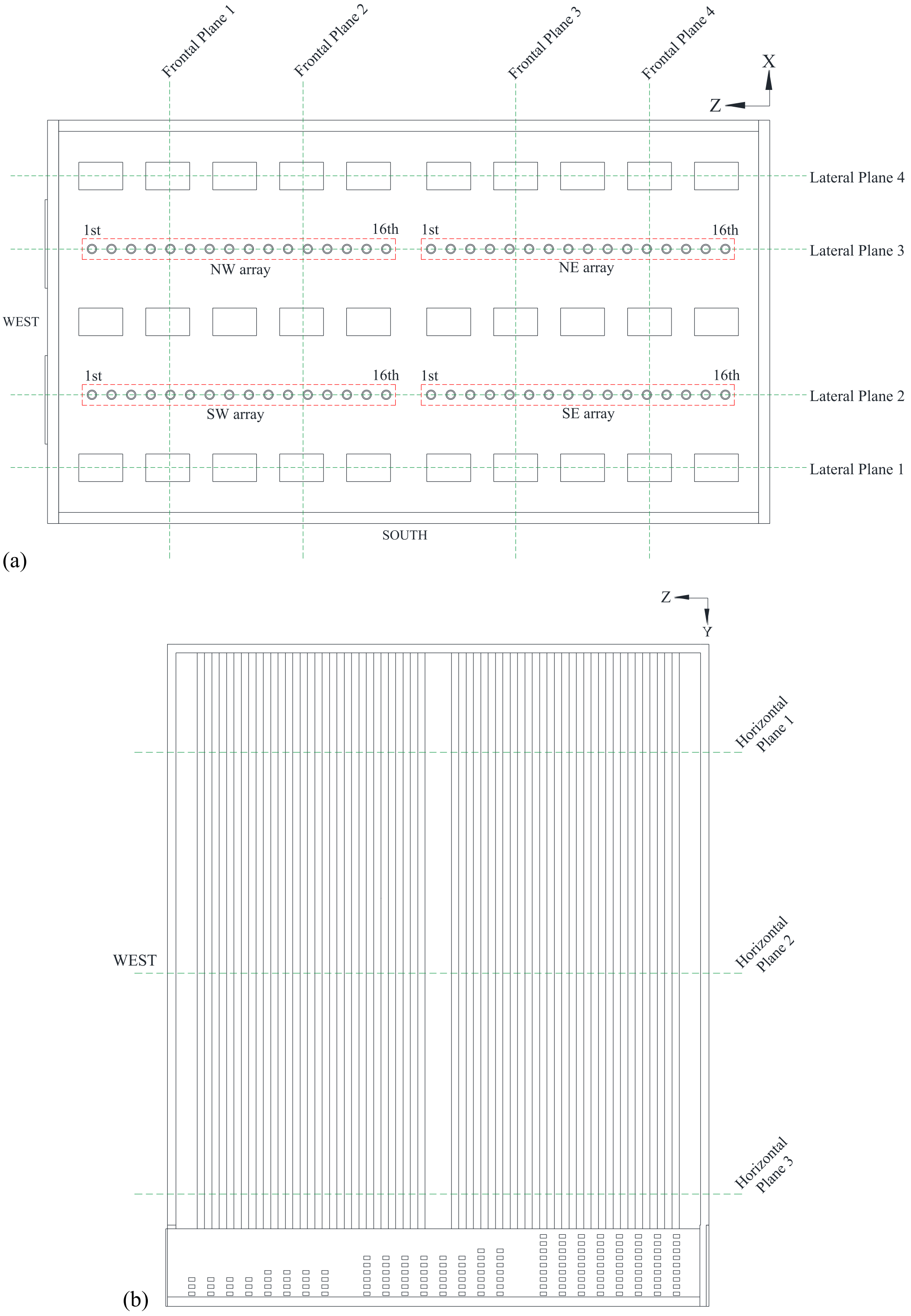

**Fig. 3.** Top and side views of the full-scale furnace; a) Top view with different frontal and lateral cutting planes indicated; b) Side view with different horizontal cutting planes indicated.

The quality of the mesh plays a significant role in the accuracy and stability of the numerical study. One of the advantages of using polyhedral mesh elements for grid construction is that highly skewed elements are prevented. High skewness can cause various numerical problems. In any CFD study, skewness angles of the elements are recommended not to exceed 80 degrees. Skewness an-gles of elements are distributed between 0 - 40 and 92 percent of the elements have skewness angles of below 5 degrees.

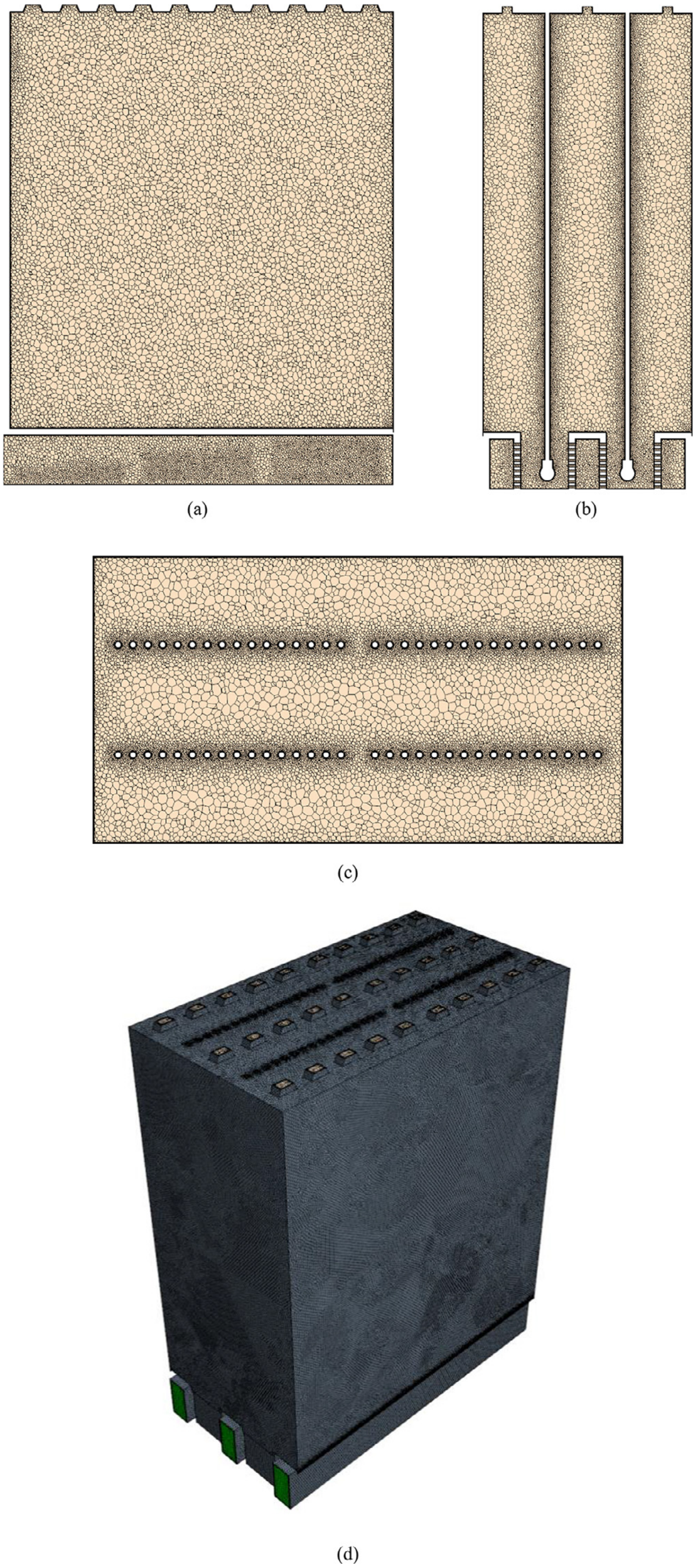

**Fig. 4.** The local and global mesh layouts of the full 3-D domain; a) Local mesh layout in the vertical middle cross-section of *-yz* plane; b) Local mesh layout in the vertical middle cross-section of *-xy* plane; c) Local mesh layout in the horizontal middle cross-section of *-xz* plane; d) Global mesh layout.

The cell quality metric algorithm is based on a hybrid of the Gauss and least-squares methods for cell gradient calculation methods. It is a function not only of the relative geometric distribution of the cell centroids of the face neighbor cells, but also of the orientation of the cell faces. Generally, flat cells with highly non-orthogonal faces have a low cell quality. A cell with a quality of 1.0 is considered perfect. A cubic cell is an example of such a perfect cell, but other polyhedral cell shapes can also have a cell quality approaching unity. A degenerate cell has a cell quality approaching zero. Cells with a face quality less than $1.0 \times 10^{-5}$ are considered to be coarse. Depending on the physics that is selected for the analysis, the cell quality of a cell can be fairly low and still provide a valid solution. However, poor cell quality is likely to affect both the robustness and accuracy of the solution. Taking into account low skewness values and low aspect ratios for constructed mesh, cell quality measure values are distributed between 0.5 and 1.

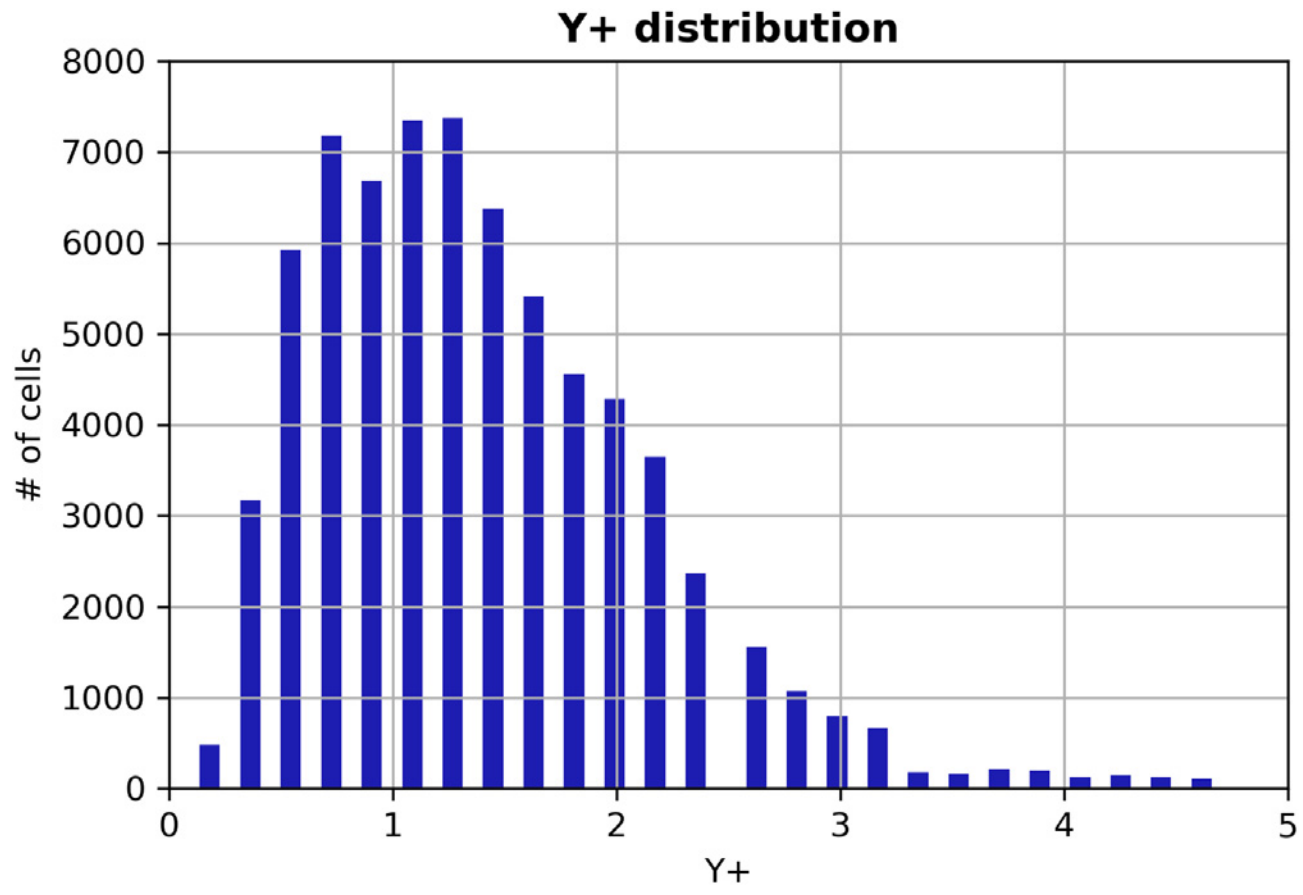

**Fig. 5.** Local cell Reynolds number, $y^+$ tests for arbitrarily chosen four reforming tube (RT) walls (one from each row) and two furnace walls namely north panel wall (NPW) and south panel wall (SPW) for a final mesh system of 16,800,000 cells.

### *3.3. Simulation set-up*

#### *3.3.1. Model coupling*

The three-dimensional (3-D) steady-state governing equations, i.e. continuity, momentum and energy equations together with species conservation equations for a compressible, reacting fluid are solved by a finite volume method (FVM) based fluid flow solver (STAR-CCM+ 2020). The Boussinesq eddy viscosity based two-equation turbulence models are used to close the Reynolds averaged Navier-Stokes equations by relating the Reynolds stress terms to the averaged velocity gradient. For steady-state simulations, each flow variable in the RANS equations is decomposed into its time-averaged quantities. This approach provides a good prediction of flow field variables with a relatively low computational cost.

For the combustion modeling, two different combustion models are evaluated. First one is the standard Eddy Break-Up (EBU) model which is a well-known combustion model with a wide range of applications. In EBU model, a global reduced chemical kinetic mechanism for premixed combustion of the fuel content is used. The second one is the Flamelet Generated Manifold (FGM) model which is an accurate and computationally inexpensive model. FGM model is a chemistry reduction method that combines the advantages of chemistry reduction and flamelet models and is suitable for most premixed/non-premixed flame applications such as gas turbines, furnaces and burners (Donini et al., 2013). FGM creates lookup tables using the detailed chemical kinetic mechanisms before the simulation starts. These lookup tables provide reaction products and mixture properties for a given range of states which reduces computational time. By using FGM with a sub-model called Turbulent Flame Speed Closure (TFC) here, a robust and accurate combustion mechanisms is utilized.

For the SMR process which occurs inside the RTs, a specialized model called Reacting Channel (RC) model is employed. RC model allows the solver to model reactions in a 1-D manner which does not take radial mixing effects into account and makes the simulations computationally efficient compared to 3-D modeling of all reacting channels in the system. The flow inside the RT is approximated as a plug flow (radial variations are ignored). This is suitable for geometries such as that of an SMR furnace, where the reacting fluid that resides inside long and relatively thin RTs exchange heat with the flue-gas outside the RTs.

A participating media radiation model called the Discrete Ordinates (DO) model is the best suited model for computing radiation problems with localized sources of heat and broad application range (Habibi et al., 2007). Thus, for radiation modeling DO model is utilized with weighted-sum-of-gray-gases radiation sub-model (Modest, 1991) which, is the preferred model in the literature as well.

Segregated flow solver is employed in the solution which solves the integral conservation equations of mass and momentum in a sequential manner. The segregated solver employs a pressure-velocity coupling algorithm where the mass conservation constraint on the velocity field is fulfilled by solving a pressure-correction equation. The Semi-Implicit Method for Pressure-Linked Equations (SIMPLE) pressure-velocity coupling scheme is well-suited for steady cases as it is the case for base model of the furnace.

#### 3.3.2. Boundary conditions

In the present simulation set up, the following boundary conditions are implemented at the boundary surfaces of the computational domain:

For firebox modeling;

- *Inlet boundaries*: Mass flow inlet boundaries are imposed at each burner entrance (opening) with a specified value of 0.195 kg/s (fuel and air mixture) with an average gauge pressure of 0.587 kg/cm$^2$ based on experimental measurements given in Table 5.
- *Outlet boundaries*: Outflow boundary conditions are imposed at three outlet sections placed at the bottom of the west plane to ensure that the flue-gas passes to convection section of the furnace.
- *Wall boundaries*: No-slip velocity boundary conditions together with a constant heat flux of -97.26 W/m$^2$ thermal conditions are defined at remaining boundaries of the computational domain with emissivity values of 0.6 (Tran et al., 2018).

In addition to the above boundary conditions, initial temperature field at ambient air temperature of 25 °C, initial pressure field as 101,325 Pa and velocity field at stagnation conditions (zero velocity) are also defined together with the turbulence kinetic energy and dissipation rate with their minimum default values.

For RC modeling;

*Specified inlet boundaries*: Although a 1-D modeling strategy is employed in the solution of turbulence-chemistry interactions inside RTs, it is required to define inlet areas, mass fractions, pressure, temperature and velocity.

*Tube wall boundaries*: No-slip velocity boundary conditions together with conjugate heat transfer thermal conditions are defined at remaining boundaries of the computational domain with emissivity values of 0.85 (Latham, 2008, Tran et al., 2018).

In addition to specified inlet boundaries, to account for the presence of a catalyst inside the RTs, the reactions are modeled in a packed bed system with Ergun correlation. Particle diameter and heat transfer factor are also specified as 0.0075 m and 1.68, respectively. Inlet velocity of the RTs are defined to be 1.45 m/s. These values can be revised according to process data.

#### 3.3.3. Computational requirements

All case simulations are solved on a 16-processor Intel Xeon® CPU E5-2690 v4 2.6 GHz machine, with 128 GB RAM per unit by using a parallel computing technology. The proposed numerical methodology uses a single programme multiple data passing model. With this technology, as the number of compute nodes increases, turnaround time for the solution will decrease in a more cost effective and robust numerical approach.

#### 3.3.4. Remarks

Although there are prior studies on the CFD modeling of SMR furnaces (Zheng et al., 2010a, Zheng et al., 2010b, Tran et al., 2017b, Lao et al., 2016, Tran et al., 2017a, Tran et al., 2018, Chen et al., 2019), majority of them do not perform a combined firebox – reforming tubes analysis. It is apparent that the reforming tube side and the furnace side interact with each other strongly and neither of them can be investigated independently. As a consequence, a good coupling between the furnace side and the reforming side is highly necessary. Furthermore, as noted by Chen et al. (2019) there is still a lack of in-depth studies on the selection of coupling methods for industrial-scale SMR furnaces. Therefore, the CFD literature is quite limited within this context. The following remarks can be made about the novelty of the computational methodology of the present study.

- The current study makes use of a comprehensive industrial data set to validate the model based on both furnace side and reforming tube side parameters.
- The present study for the first time provides most appropriate – optimum choice of the RANS equations-based turbulence models to be accompanied with combustion and radiation models both i.e. choice of appropriate turbulence-combustion-radiation models are achieved here.
- Grid independency of the flow calculations is achieved by using a comparatively less number of fully unstructured polyhedral and prism layer based cells, which leads to less computational cost without giving compromise on the numerical accuracy and stability.
- The combustion model used in the present study, namely Flame Generated Manifold (FGM), is applied for the first time for SMR furnace simulations.
- Conjugate heat transfer boundary conditions at the reforming tube walls are applied rather than solving it with user-defined temperature or heat flux profile implementation which makes it suitable for cases in which there is no reliable reforming tube wall temperature data.

### 3.4. Mesh sensitivity analysis

The effect of the mesh resolution on the present 3-D flow is tested using the Realizable k-$\varepsilon$ turbulence model which is coupled with Flamelet Generated Manifold (FGM) model, Discrete-Ordinates (DO) model with weighted-sum-of-gray-gases (WSGGM) radiation sub-model and a specialized Reacting Channel (RC) model for four different mesh resolutions containing nonuniformly distributed cells from 12.88 million to 18.55 million. The Realizable k-$\varepsilon$ turbulence model is chosen as a reference turbulence model here as it has been successfully utilized for steady combustion applications of other large industrial systems (Stefanidis et al., 2006, Tutar, 1998) and also for reverse flow dominated aerodynamic flows (Zečević and Bolf, 2020). As previously observed from the local cell Reynolds number, $y^+$ evolution along the arbitrary chosen RT walls and NPW and SPW in Fig. 5, the mesh resolution of 16,800,000 cells of the 3-D mesh resolution represents well for $y+$ values, which are as low as 0.1. A quantitative analysis of temperature profile along the wall distance in the $y$-direction for each mesh resolution (not shown here) suggests that the mesh systems which contain less than 16,800,000 cells reproduce very distinctive and scattered temperature profiles compared to finer mesh systems and the difference between mesh resolutions of 16,800,000 and 18,500,000 cells is very low and both mesh systems reproduce very similar temperature profiles. On the other hand, a further quantitative comparison in terms of the calculated area-weighted average outlet temperature for each mesh system clearly demonstrates the effect of mesh resolution on the temperature profiles as seen in Fig. 6. Only minor changes in the calculated values are observed when more than 16,800,000 million cells are employed (the relative error in the calculated value is around 0.02 % when the mesh resolution is increased from 16,800,000 cells to 18,500,000 cells by $\approx$ 10 %). Therefore, a mesh system of 16,800,000 cells is assumed to be a good compromise between accuracy and computational time for the present 3-D case and is chosen as an optimum, final mesh system for further analysis in the subsequent sections.

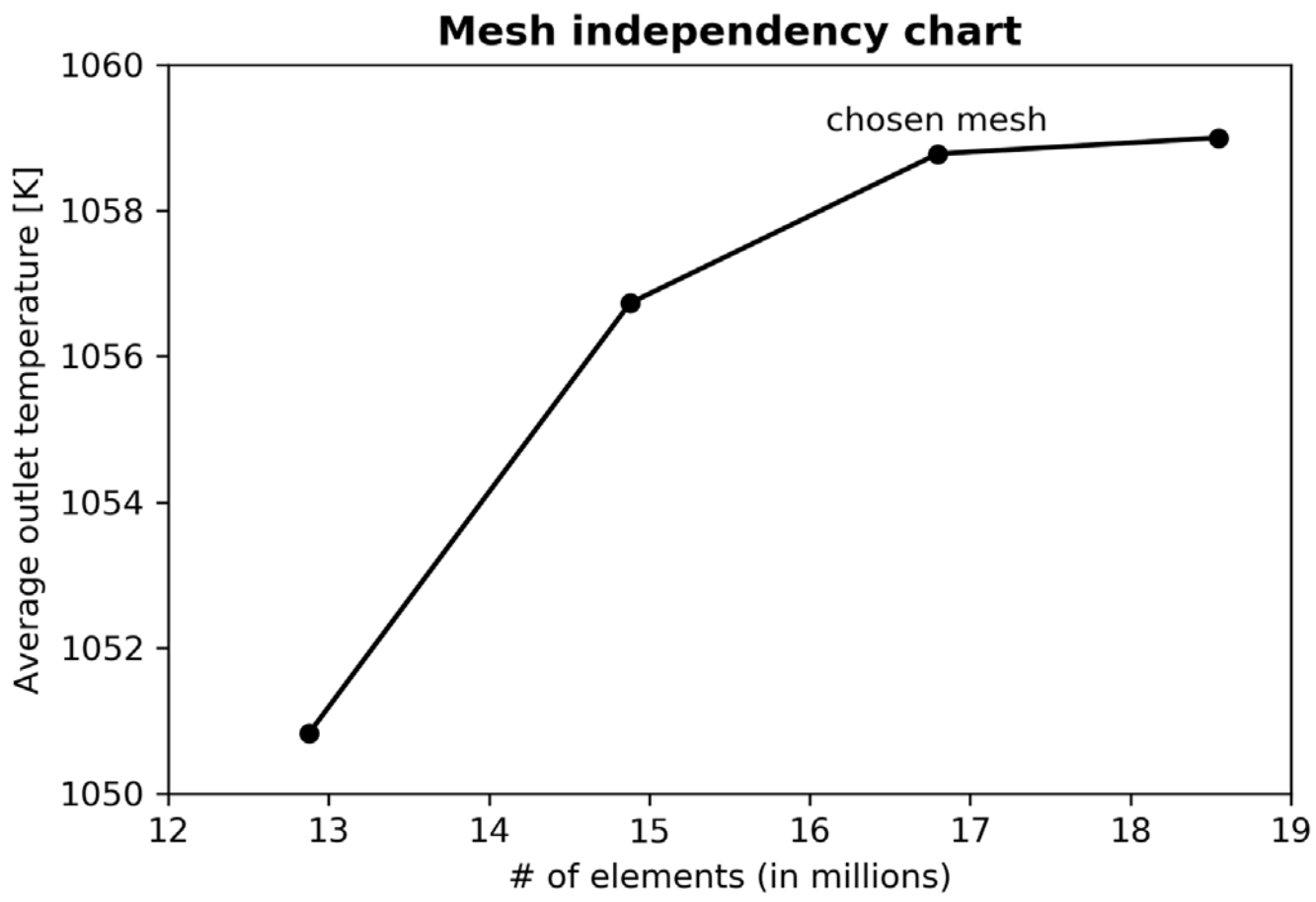

**Fig. 6.** Mesh sensitivity analysis for the present 3-D full-scale 3-D simulations performed with the Realizable k-$\varepsilon$ turbulence model which is coupled with Flamelet Generated Manifold (FGM) model, Discrete-Ordinates (DO) model with Weighted-sum-of-gray-gases (WSGGM) radiation sub-model and 1-D plug flow Reacting Channel (RC) model.

## 4. Results and discussions

The suggested coupled modeling and simulation framework presents different simulation cases which are run in accordance with the available boundary, operating and other physical flow conditions of an industrial SMR furnace. The results obtained from each simulation case (SC) are comparatively analyzed in a quantitative manner to demonstrate the effect of choice of coupled modeling and the capability of full 3-D CFD modeling approach on the prediction of the physical processes involving turbulence, combustion, radiation and the catalytic reactions. Thorough qualitative and quantitative analysis of the firebox at different planes and the reforming tubes are further accomplished with the chosen coupled modeling approach.

### *4.1. The analysis of flow and heat in the firebox*

#### *4.1.1. Average temperatures*

Having completed mesh independency tests, modeling effects on the present flow are investigated with the use of different coupled modeling approaches. Each 3-D simulation case (SC) is run under the same assumptions and modeling set ups with a selected mesh resolution of 16,800,000 cells to make a direct comparison. Quantitative comparisons between SC results and experimental data (Table 5) are made to determine the relative performance of each SC on the governing flow features and heat distributions of the furnace.

The results are initially obtained in terms of line average temperature at $y$ = 10 m of NPW, $T_{TS\text{-}1}$, $_{TS\text{-}2}$, and SPW, $T_{TS\text{-}3}$, $_{TS\text{-}4}$, area-weighted average temperatures of outlet manifolds of reforming tubes $T_{TS\text{-}7}$, $T_{TS\text{-}8}$, $T_{TS\text{-}9}$, and $T_{TS\text{-}10}$, and finally area-weighted average temperature of two outlets of the combustion gases, $T_{TS\text{-}11}$, $_{TS\text{-}12}$, for three different simulations which are performed under the same dimensional modeling i.e. 3-D with full flow domains and their values are summarized comparatively in Table 6. As can be identified from Table 6, SC-2 reproduces higher average temperature predictions for each zone and these values are in a better correspondence with the experimental data. Moreover, the methane conversion (%) is closer to experimental data for SC-2. SC-2 also have relatively low computational cost in comparison with the other models except SC-1 which uses a more simplified turbulence model.

The relative errors between the SC-2 and experimental data for temperatures are at most 4.77 %, while for methane conversion (%) it is 16.80 %. This relatively high error for methane conversion (%) can be attributed to the kinetic constants of the 1D turbulence-chemistry model that is used to model the reforming phenomenon. Nevertheless, this does not significantly affect the furnace side flow and temperature fields qualitatively and overall correspondence of the results with the experimental data is acceptable.

#### *4.1.2. Velocity fields*

The present simulation case (SC2) provides similar velocity fields at different vertical –xy cutting planes i.e. frontal planes (Fig. 7 (a) to (d)) with the prediction of large recirculation zones which occur in the upper left and right parts of each burner lane close to the flame region due to entrainment of the jet (the burners can be considered to be jets here as the air-fuel mixture is entered into the flow domain through them at higher velocity) into a confined space between the rows of the tubes. The flow visualization method used in Fig. 7 (a) to (d) is the line integral convolution method (LIC). LIC method outputs a dense streamline plot that represents all 2D features of the input vector field. The number and extent of the recirculation zones, which significantly affect the flue-gas temperature distributions in the firebox, are found to be almost the same at each FP where the flame patterns also exhibit very similar evolution characteristics with the one which has a tendency of bending towards the north side RT row in the center burner lane, as seen in Fig. 7 (a) to (d).

The maximum flue-gas velocity vectors are found to be in the outlet regions inside the evacuation tunnels, and the highest values of the maximum velocity is obtained with a value of 10.24 m/s at the FP-1 which is the closest plane to the west panel wall (WPW) of the furnace. The flue-gas velocity vectors on the other hand become very small in the vicinity of the RT walls and in the center of the firebox (large dead flow zones). Additionally, it is seen that there is a flame impingement on the north side reforming tube walls, as observed in Fig 7 (a) and (d). The impact occurs at a height which is approximately 2.20 m down from the burners.

The downward movement of the flue-gas between the RTs and the walls of the furnace following the combustion of the fuel with the air in the burners are also examined along the cross-section of different –*yz* cutting planes i.e. lateral planes, as seen in Fig. 8 (a) to (d). Due to the movement of the flue-gas towards the furnace walls and/or RTs with the existence of developing horizontal velocity vector field, there is no perfect symmetry in the flow field with respect to the vertical middle cross-section of -*yz* cutting plane. No discernible recirculation zones between each reforming tube can be found due to narrow spacing. Moreover, no large recirculation zones are observed at the lateral planes cutting through the burners, unlike the ones previously observed at the frontal planes of the furnace. The flue-gas movement towards the outlets of the fur-nace with the highest local velocity vectors are clearly identified at LP-1 and LP-4, as seen in Fig. 8 (a) and (d) respectively.

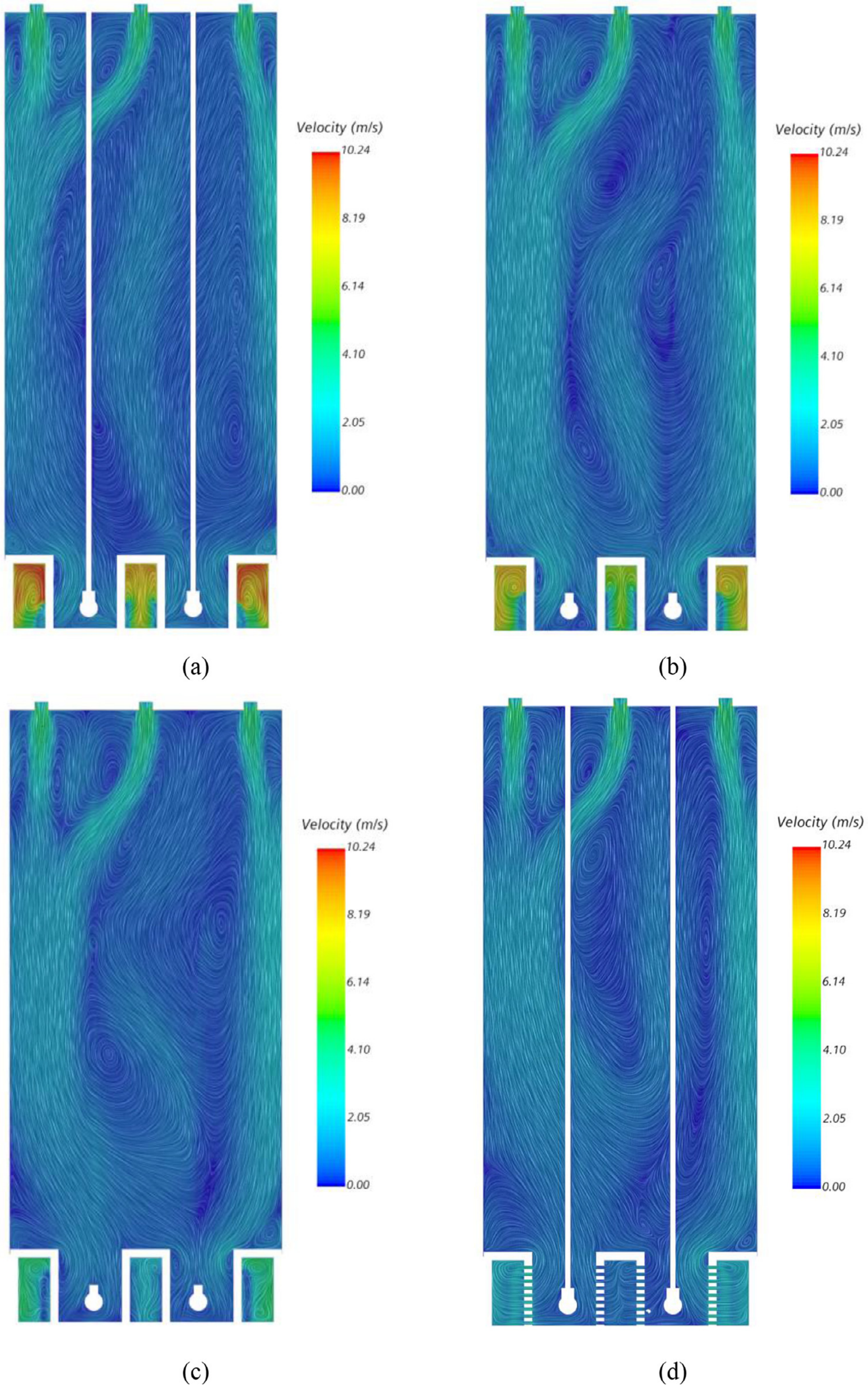

**Fig. 7.** Representation of the computed velocity vector fields using line integral convolution (LIC) at different vertical –*xy* cutting planes. Left: NPW and right: SPW; a) Frontal Plane -1 (FP-1); b) FP-2; c) FP-3; d) FP-4.

The principal downward movement of the flue-gas between the rows of the RTs and the walls of the furnace on the other hand is further investigated using 3-D surface warped vector fields represented at different horizontal planes (HPs) of varying heights, as seen in Fig. 9 (a) to (c). The simulation case predicts almost the identical velocity distributions at all HPs in the downstream direction towards the outlet sections except those at HP-1 where near-burner velocity predictions result in higher velocity vectors which locally change in the north side burner lane. The similarity of flow patterns suggests that the recirculation and the downward motion of the flue-gas be successfully estimated by the present full 3-D CFD model. There is no significant cross-flow observed between the burner lanes due to the extension of the reforming tubes along the whole height of the firebox and the existence of large recirculation zones are accompanied with very low magnitude velocity vector fields at both sides of the RTs.

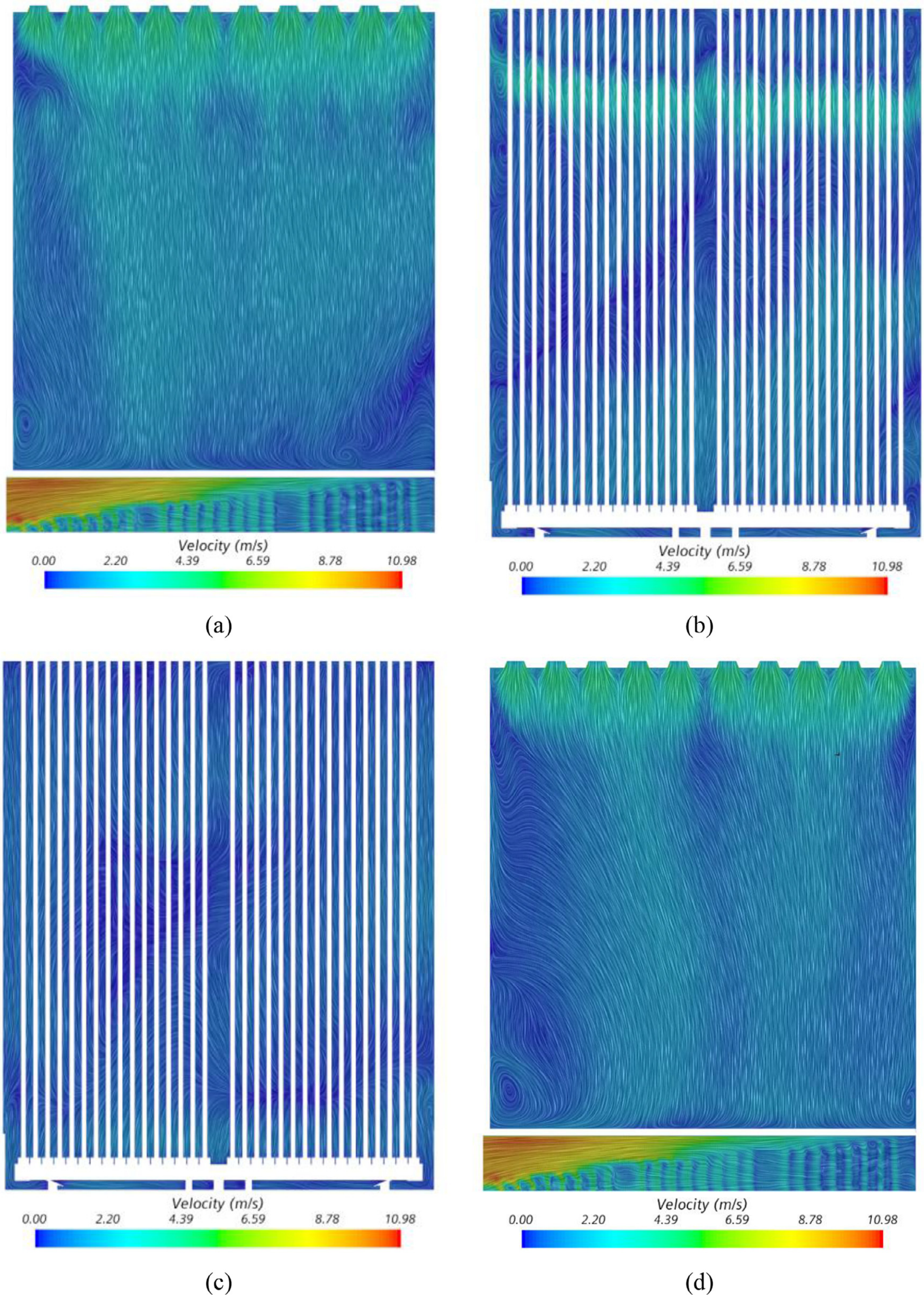

**Fig. 8.** Representation of the computed velocity vector fields using line integral convolution (LIC) at different vertical –*yz* cutting planes. Left: WPW and right: EPW; a) Left Plane -1 (LP-1); b) LP-2; c) LP-3; d) LP-4.

Three-dimensional (3-D) flue-gas velocity field and the streamlines colored by the velocity magnitude are also computed and presented in the full furnace as seen in Fig. 10. The starting and ending points of the streamlines are marked for easy monitoring. 3-D asymmetrical downward movement of flue-gas with relatively higher velocity vectors following the combustion of the fuel with the air are more discernible in Fig. 10. The strong interaction of flue-gas patterns between the center burner lane and the north side burner lane is observed. This leads to an uneven distribution of the flue-gas temperature in the firebox which may turn in non-symmetrical temperature evolution around the RTs. The exit of the flue-gas through the outlet sections with increasing velocity vectors in association with local vortical structures is also identified. The highest velocity magnitudes around 8.0 – 10.0 m/s are observed at the outlets where the pressures are the lowest in the domain.

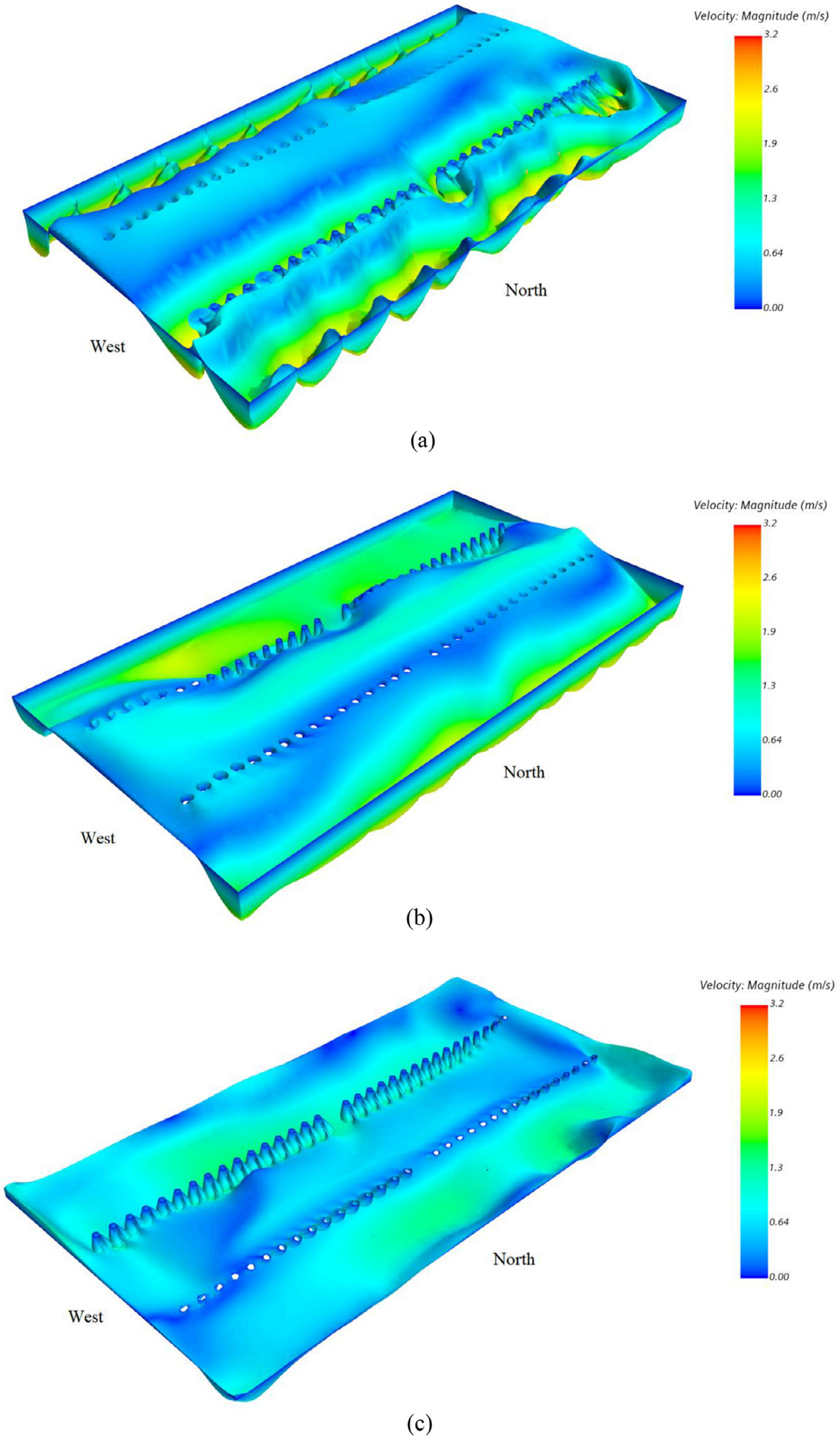

**Fig. 9.** Representation of the computed velocity vector fields using scalar warping at different horizontal –*xz* cutting planes of varying heights; a) Horizontal Plane -1 (HP-1); b) HP-2; c) HP-3.

**Table 6**
Calculated average temperatures for different simulation cases which are run for final, optimum mesh resolution containing 16,800,000 cells for 3-D flow-domain ($\varepsilon$ represents the relative error).

| Simulation Cases (SCs) | $T_{TS\text{-}1,\,TS\text{-}2}$ [K] ($\varepsilon$ %) | $T_{TS\text{-}3,\,TS\text{-}4}$ [K] ($\varepsilon$ %) | $T_{TS\text{-}7}$ [K] ($\varepsilon$ %) | $T_{TS\text{-}8}$ [K] ($\varepsilon$ %) | $T_{TS\text{-}9}$ [K] ($\varepsilon$ %) | $T_{TS\text{-}10}$ [K] ($\varepsilon$ %) | $T_{TS\text{-}11,\,TS\text{-}12}$ [K] ($\varepsilon$ %) | Methane conversion % ($\varepsilon$ %) | Computation time [h] |
|---|---|---|---|---|---|---|---|---|---|
| SC-1: Standard k-$\varepsilon$ FGM/TFC – DO – RC models | 1303.6 (3.56%) | 1286.9 (4.25%) | 991.2 (3.51%) | 972.4 (3.12%) | 971.1 (5.24%) | 964.7 (4.52%) | 1053.2 (1.01%) | 84.1 (18.45%) | 26.70 |
| SC-2: Realizable k-$\varepsilon$ FGM/TFC – DO – RC models | 1320.05 (2.32%) | 1303.2 (2.93%) | 995.8 (3.06%) | 980.8 (2.28%) | 975.9 (4.77%) | 964.5 (4.53%) | 1057.8 (0.57%) | 82.9 (16.80%) | 28.64 |
| SC-3: Standard k-FGM/TFC – DO – RC models | 1307.6 (3.25%) | 1292.6 (3.73%) | 994.2 (3.22%) | 974.2 (2.94%) | 973.9 (4.97%) | 960.8 (4.90%) | 1056.8 (0.67%) | 82.6 (16.34%) | 30.61 |
| SC-4: Standard k-$\omega$ SST FGM/TFC – DO – RC models | 1302.43 (3.63%) | 1274.3 (5.09%) | 993.4 (3.29%) | 976.6 (2.70%) | 972.3 (5.12%) | 958.8 (5.10%) | 1057.0 (0.65%) | 82.9 (16.80%) | 31.21 |
| SC-5: Realizable k-$\varepsilon$ EBU – DO – RC models | 1318.1 (2.47%) | 1299.9 (3.18%) | 994.1 (3.23%) | 979.2 (2.44%) | 971.7 (5.19%) | 962.2 (4.77%) | 1084.6 (1.95%) | 86.2 (21.41%) | 38.56 |
| Experimental Data | 1351.4 | 1342.6 | 1027.2 | 1003.7 | 1024.8 | 1010.3 | 1063.9 | 71 | NA |

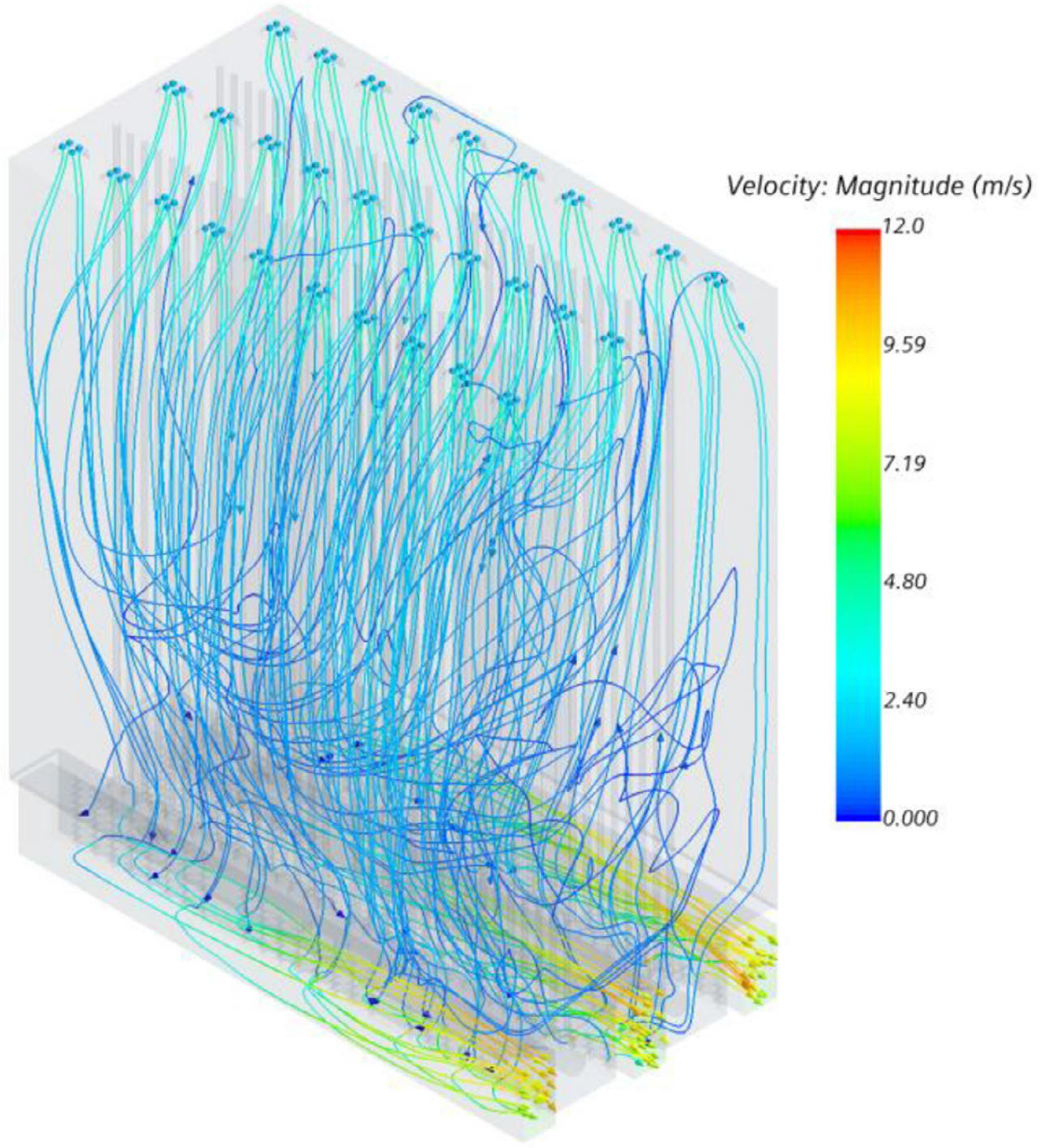

**Fig. 10.** Representation of the computed 3-D velocity fields in the full-scale furnace using streamlines colored by velocity magnitude.

#### *4.1.3. Temperature fields*

Flue-gas velocity field directly affects the flue-gas temperature distribution in the firebox as seen in Figs. 11 to 13. A maximum temperature value is computed in the flame front region from where the highest temperature gradient towards the reforming tubes is observed. The flue-gas temperature is therefore much higher in the close proximity of the burners and gradually decreases downwards.

Fig. 11 shows the temperature fields observed at different frontal planes i.e. FPs from the west side to the east side of the furnace (see Fig. 3). The higher temperatures are observed in the close proximity of the north and south panel walls where the flue-gas has a short residence time as expected. The center burner lane of the furnace seems to have lower temperatures for all FPs. A considerable radiative heat flux to the reforming tubes along their lengths leads to relatively lower temperature field in all fluid zones and this temperature reduction is more discernible in the center burner lane due to larger heat-absorbing surface area of the reforming tubes, re-circulations and the flame bending at the middle burner lane. Therefore, lower temperature values are obtained in the center burner lane compared to the left-hand and right-hand side burner lanes. Indeed, asymmetric temperature distribution with respect to the centerline of the firebox appears due to asymmetrical downward movement of flue-gas associated with the flame bending.

Fig. 12 (a) to (e) shows the temperature fields observed at different lateral planes i.e. LPs from the south side to the north side of the furnace (see Fig. 3). As expected, high temperature gradients are present at upper-heights of the furnace, near the flames. When compared, slightly higher temperatures are observed for the north side of the furnace (FP-4). It is also seen that the west side of the furnace has lower temperatures compared to east side. The temperatures decrease downwards, towards the outlets i.e. flue gas tunnel exits.

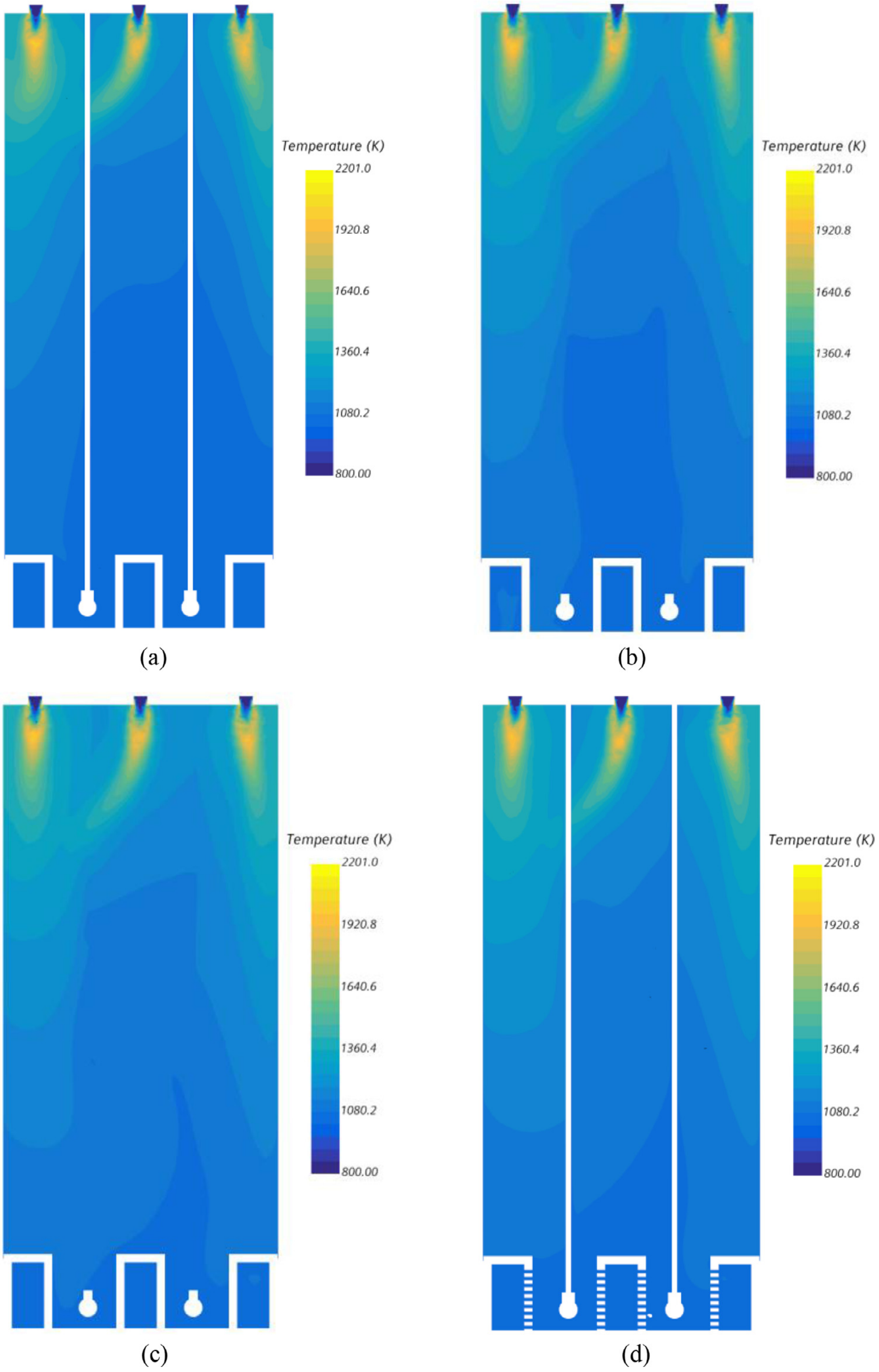

**Fig. 11.** Representation of computed static temperature fields at different vertical –*xy* cutting planes (frontal planes). Left: NPW and right: SPW; a) FP-1; b) FP-2; c) FP-3; d) FP-4.

On the other hand, Fig. 13 (a) to (e) shows the temperature fields observed at different horizontal planes i.e. HPs from the top to the bottom of the furnace (see Fig. 3). Higher temperature values with more asymmetrical temperature distribution are obtained in the order of height for all horizontal flow domains. It is seen that the asymmetric temperature distributions and high gradients tend to diminish in the downward direction. Therefore, the uni-formity in the temperature distributions is more discernible in the lower heights of the section. This is due to more pronounced mixing effects of heat diffusion and convection mechanism, and hence, heat recovery that is attained in the lower height of the firebox in the center burner lane. However, at all heights north and south (especially the north) sides of the furnace have higher temperatures compared to center of the burner, as stated earlier. Additionally, at all heights, west side of the furnace has lower temperatures compared to all other regions of the furnace.

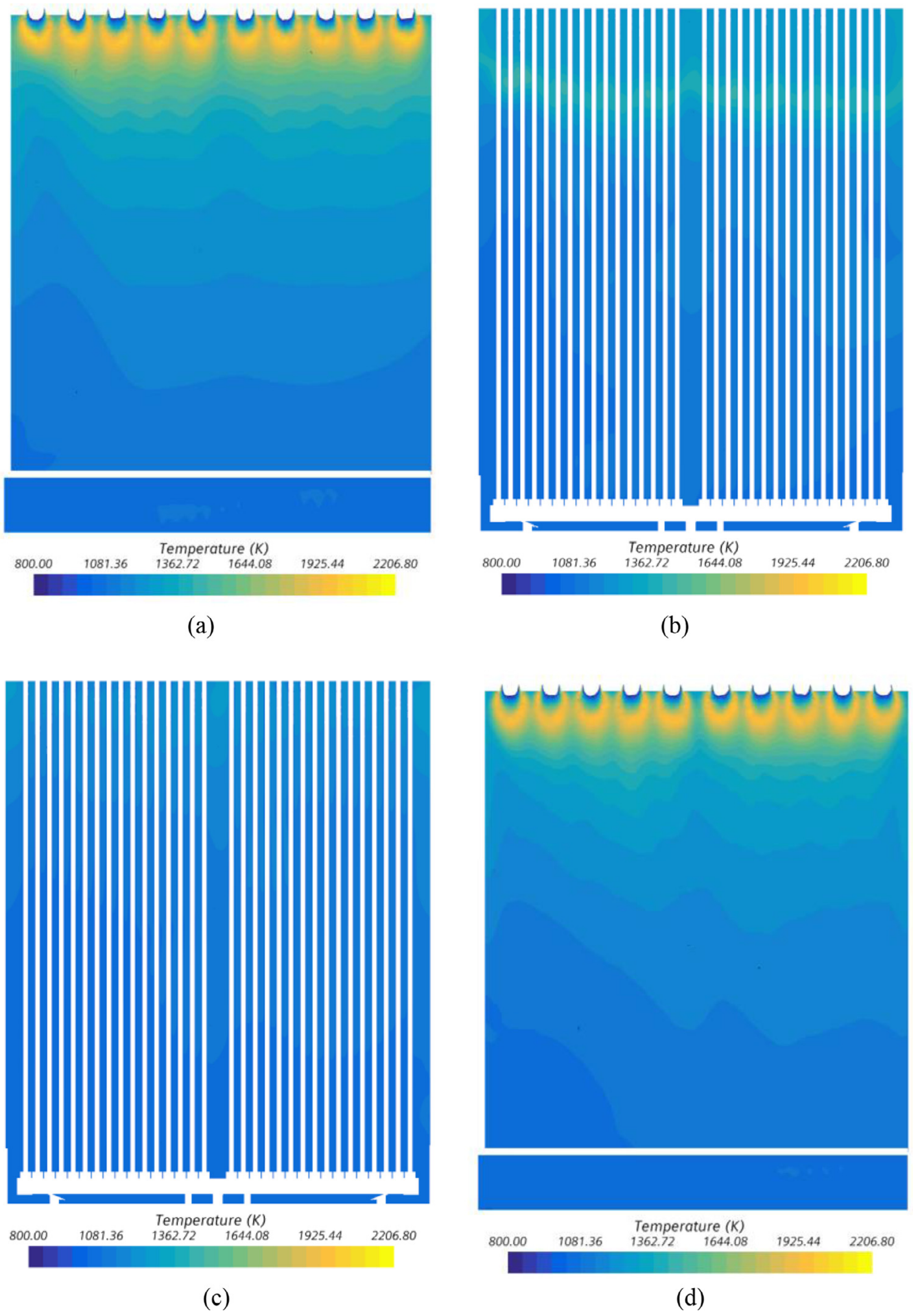

**Fig. 12.** Representation of computed static temperature fields at different vertical –*yz* cutting planes (lateral planes). Left: NPW and right: SPW; a) LP-1; b) LP-2; c) LP-3; d) LP-4.

The maximum temperature and the flame height predicted by the current full 3-D model simulation are 2207 K and 0.95 m, respectively. The lowest temperature values are computed around the outlets of the furnace (~1058 K).

### 4.1.4. Temperature profiles

Temperature profile measurements in terms of *z* direction-based scattered surface static temperature values (data) along the heights of the furnace walls i.e. NPW and SPW are made as seen in Fig. 14 (a) and (b). Almost identical temperature evolution i.e. temperature increment is predicted for both wall surfaces of the furnace. At the lower heights of the walls, lower temperature values with slight increase in the order of height are predicted in comparison with those higher values obtained at the higher heights of the walls after the height of $\approx 8$ m where the increase in the temperature profile is significant due to very high radiative heat transfer to the walls. The highest average wall temperature value is computed at the end of the wall surface.

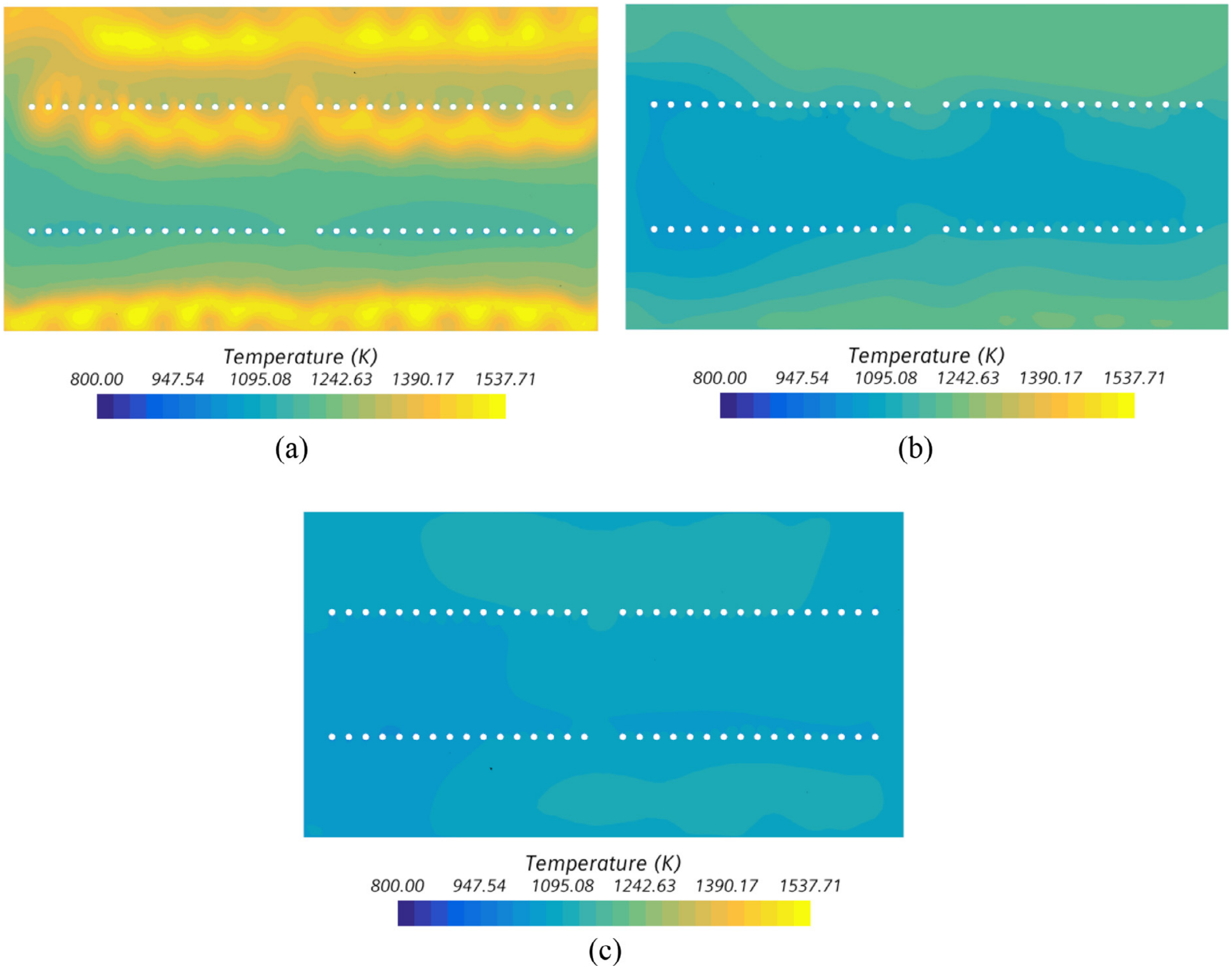

**Fig. 13.** Representation of computed static temperature fields at different horizontal –*xz* cutting planes. Left: WPW and right: EPW; a) HP-1; b) HP-2; c) HP-3.

The computed scattered surface static temperature values are later compared with the available spatial–average experimental data which are retrieved from TS-1, TS-2, TS-3 and TS-4 placed at the height of $y = 10$ m on both NPW and SPW of the furnace and from TS-5 and TS-6 placed at the height of $y = 7$ m on NPW of the furnace (see Fig. 2). Temperature sensor measurements, which are made at midnight for time duration of 80 days, represent a standard deviation of 3–4 %. When a comparison is made, it is seen that at the same furnace height of $y = 9.98$ m, the computed spatial average values for SPW and NPW are 1342.6 K and 1351.4 K, respectively and these values show a very good correlation with the spatial and time average experimental data. On the other hand, at the height of $y = 7.5$ m, computed spatial average values seem to overshoot the sensor readings. This might be due to flames in the central burners directing towards NPW and resulting in higher average temperatures at the middle-heights of the furnace. The estimated absolute errors between the average experimental data and the computed average values are found to be 1.72 % and 2.76 % for the NPW and SPW at $y = 10$ m and 4.65 % for the NPW at $y = 7$ m, respectively.

One can notice that the measurements from the present SMR furnace operation are deviated from the average values since the operation of the furnace during the 80 days of measurements are quasi-steady due to operational process changes (Tables 5 and 6). Therefore, the use of dispersed temperature input data may lead to a high error margin of the computed wall temperature values in the present numerical simulations. On the other hand, in the light of the given graphical representation of quantitative comparison of experimental and numerical data in terms of temperature distributions, it can be said that the computed wall temperature profiles demonstrate a good correlation with the spatial and time average industrial data and the error margin is within the acceptable range when compared with those of above referenced computational studies (Tran et al., 2017b, Fang et al., 2017).

### 4.2. The analysis of flow and heat in the reforming tubes

#### 4.2.1. Temperature and heat flux fields

In a typical SMR furnace, 50% of the heat generated by the combustion of the fuel is absorbed by the process gas through the reformer tubes (Zečević and Bolf, 2020). The computed 3-D heat flux and temperature distributions at the outer wall surfaces of all RTs in the furnace are presented in Fig. 15 (a) and (b), respectively. At the upper parts of the reforming tube walls, where extremely high radiative heat transfer occurs between the flue-gas and the RTs, much higher heat flux values are found compared to those of lower parts of the RTs as seen in Fig. 15 (a). A very high rate of endothermic SMR process associated with this very high radiative heat transfer on the other hand leads to lower outer wall temperatures at the upper parts of the RTs as expected (Fig. 15 (b)). Almost identical temperature and heat flux variations along each RT length are observed in each row.

#### 4.2.2. Heat maps

The computed heat maps of area-weighted average heat flux values across each reforming tube wall and of area-weighted average outer reforming tube wall temperature values are presented in Fig 16 (a) and (b), respectively. The heat maps are provided for four different RT arrays namely north-east (NE) RT array, north-west (NW) RT array, south-east (SE) RT array and south-west (SW) RT array. Non-uniform temperature and heat flux distributions are reproduced in each RT array due to the non-uniform temperature distribution of the flue-gas along the width of the furnace as previously observed in Fig. 13. This has a detrimental effect in furnace balancing (and performance) as noted by Farnell and Cotton (2000). Additionally, slightly higher heat flux and temperature values are predicted for the RTs in the north side RT row i.e. NE and NW RT arrays compared to those corresponding values of the south side RT row i.e. SE and SW RT arrays. This is probably due to effect of asymmetrical movement of the flame in the center burner lane and hence the flue-gas towards the RTs in the north side RT row. Nevertheless, the average values computed by the proposed full model case for each reforming tube are consistent with each other and the full 3-D modeling approach gives an insight for better analysis of furnace balancing in terms of heat maps of the furnace. The area-weighted average tube wall temperatures of NE, NW, SE and SW arrays are 975.7 K, 967.1 K, 954.6 K and 943.9 K, respectively. The difference between maximum and minimum area-weighted average tube wall temperatures for NE, NW, SE and SW arrays are 9.12 K, 26.62 K, 9.68 K and 27.14 K, respectively. This clearly shows that the non-uniformity effects are more discernible for the west side tube arrays. Increasing the tube wall temperatures by 38 K can decrease the tube lifetime from 10 years to 1.4 years (Zečević and Bolf, 2020). Therefore, an extra effort has to be shown to balance the tube wall temperatures to achieve safe plant operation and reduce operation costs.

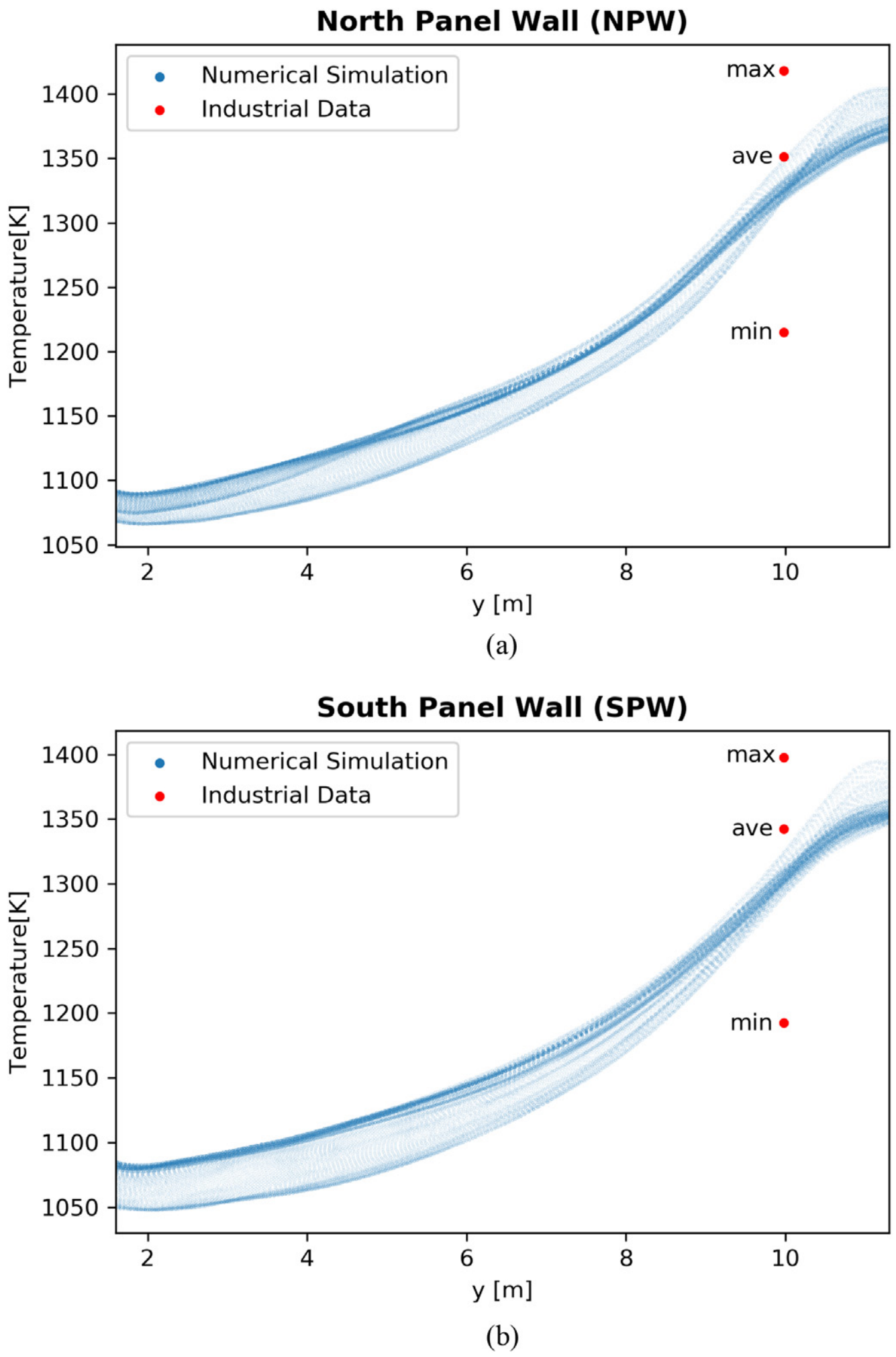

**Fig. 14.** Computed scattered surface static temperature profiles on the NPW and SPW along the height. The space and time averaged experimental data obtained for two different length points of the same height of $y$ = 9.98 m on the panel walls are also included for a comparison.

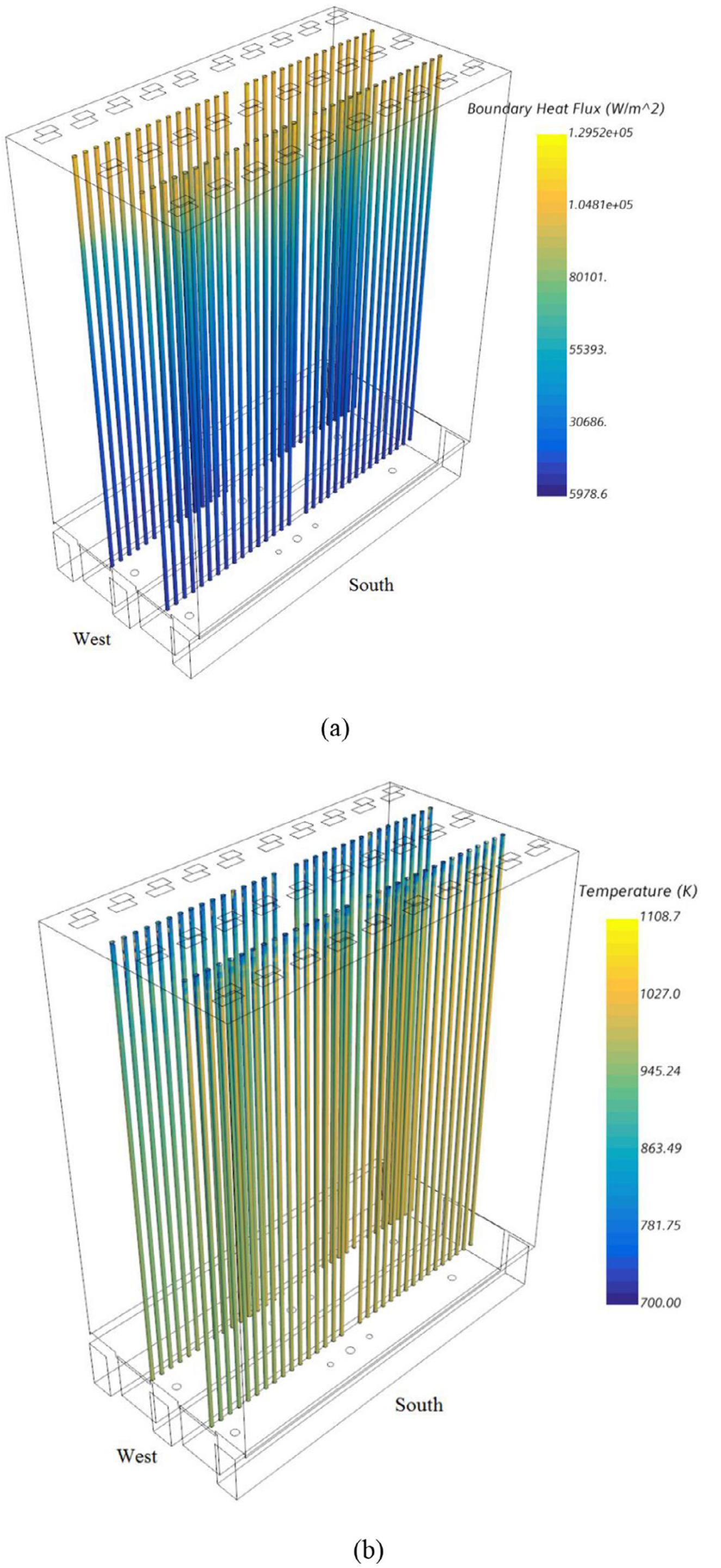

**Fig. 15.** Representation of the computed 3-D static temperature and heat flux fields at the outer RT walls.

### 4.2.3. *Temperature and heat flux profiles*

Significant amount of heat energy is extracted from the flue-gas in the upper flow zone of the furnace where the combustion of fuel takes place and is directly transferred to the reforming tube to meet the heat requirement of endothermic SMR pro-

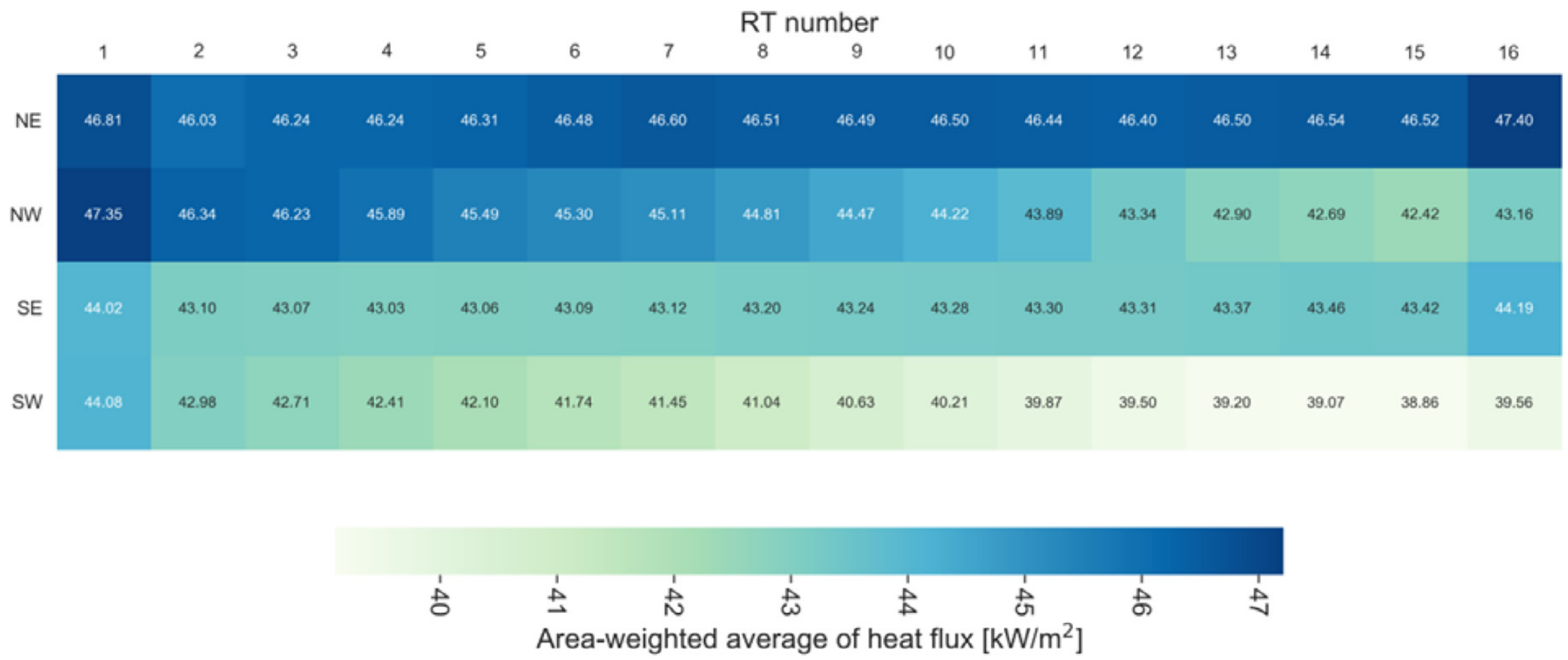

(a)

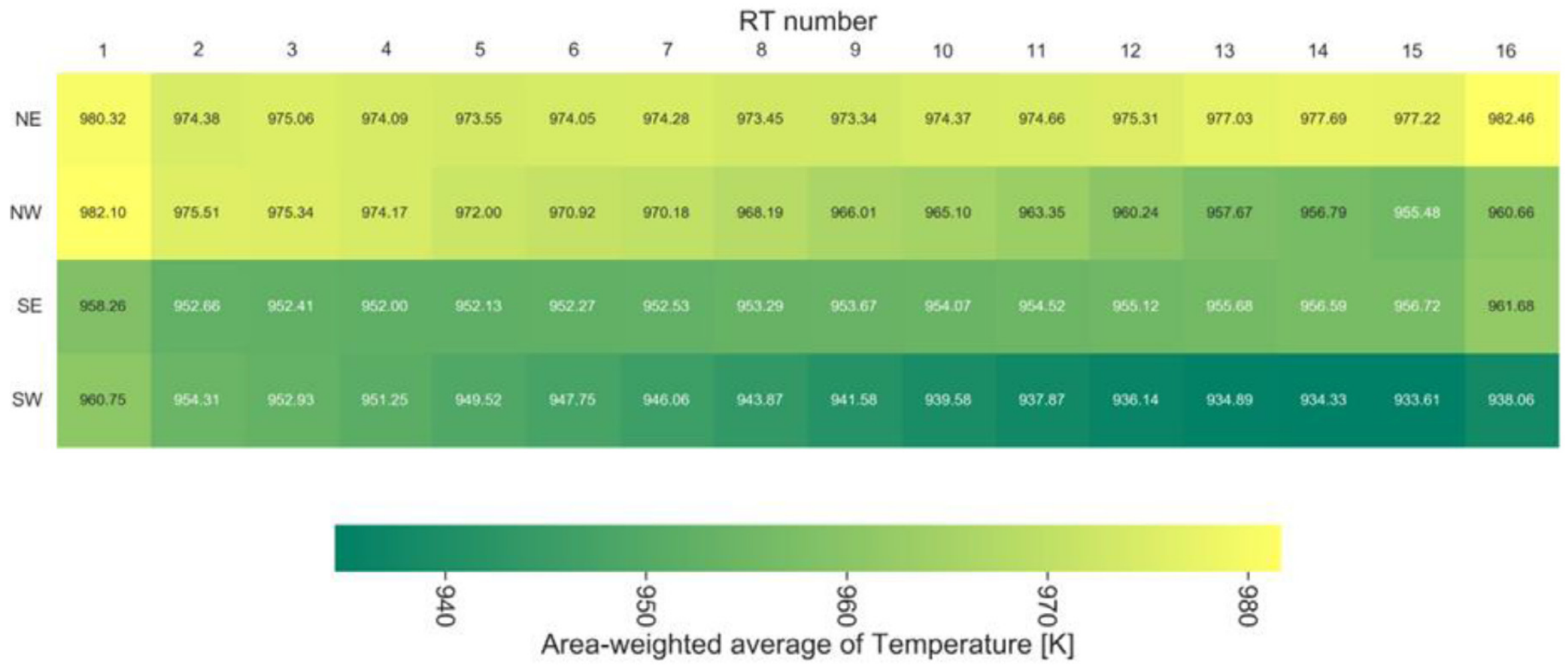

(b)

**Fig. 16.** Representation of the heat-maps of the area-weighted average heat fluxes and static temperatures of the reforming tube (RT) walls; a) Heat flux; b) Static temperature. The position and numbering of each RT according to the top view of the furnace model is depicted previously in Fig. 3 (a).

cess in each reforming tube. The heat flux distribution, which exhibits extremely high and varying/fluctuating characteristics at the upper height (beginning) of the reforming tube due to very high spatially varying flue-gas temperature, suggests that this heat requirement continues to be met along the length of the reforming tube as shown in Fig. 17. The radiative heat transfer gradually decreases towards the end of the reforming tube where the heat flux across the reforming tube wall also decreases. It is noticed that in the upper heights of the reforming tubes, the simulation case predicts slightly higher heat flux values for the reforming tubes in the NE and NW RT arrays compared to those obtained for the reforming tubes in the SE and SW RT arrays. Reaction and average outer wall temperatures for each reforming tube exhibit very similar evolution i.e. gradual increase in the downward direction as expected and reach a finite asymptote value of 929.5 K and 960.3 K, respectively for each RT array at the exit. The difference between both temperature profiles is highly discernible and is gradually alleviated as the exit of the RTs is approached. This causes a temperature drop on reaction and RT wall temperatures. It is noteworthy here that almost identical temperature and heat flux evolution

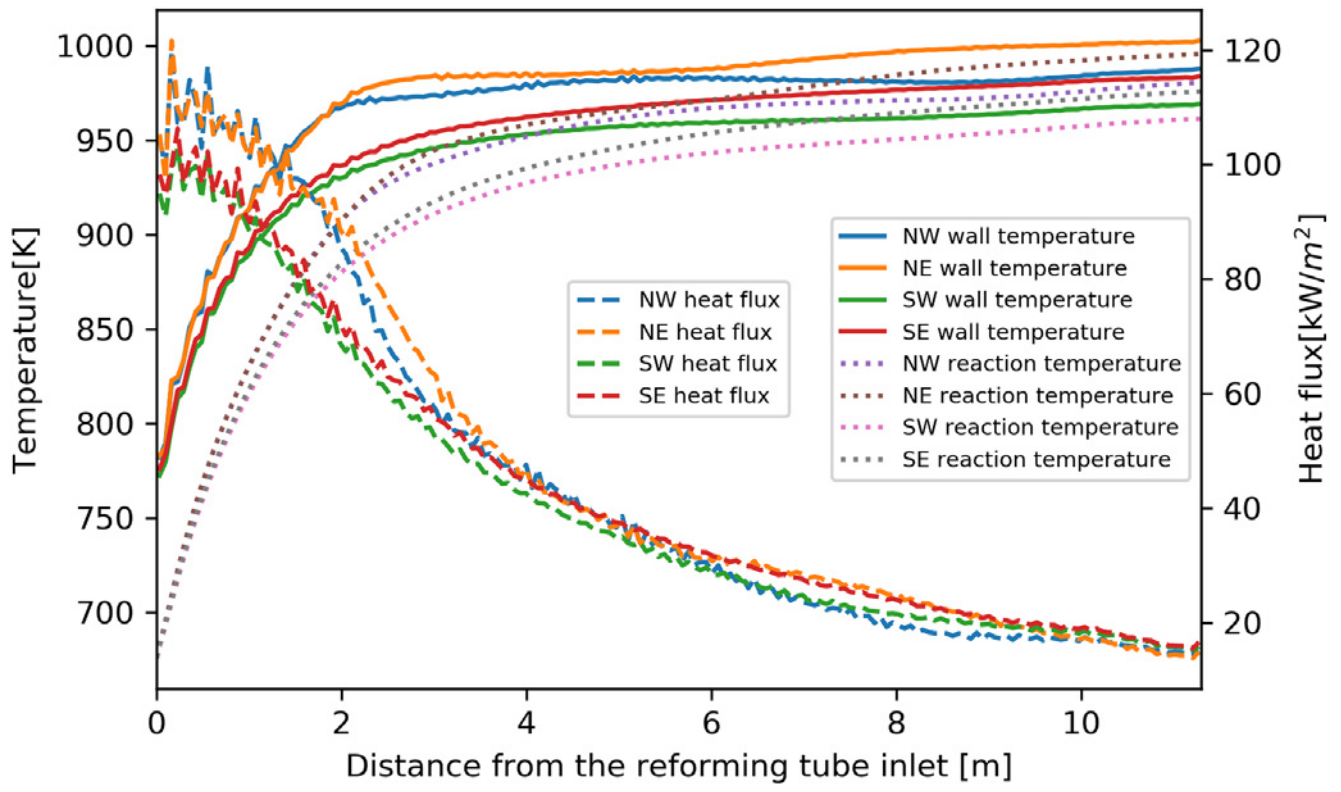

**Fig. 17.** Computed circumferential-weighted average profiles of outer reforming tube wall temperature and heat flux at the reforming tube wall together with the profiles of the RT reaction temperature along the RTs in each row.

profiles are reproduced for each RT array with a slight difference along the entrance length.

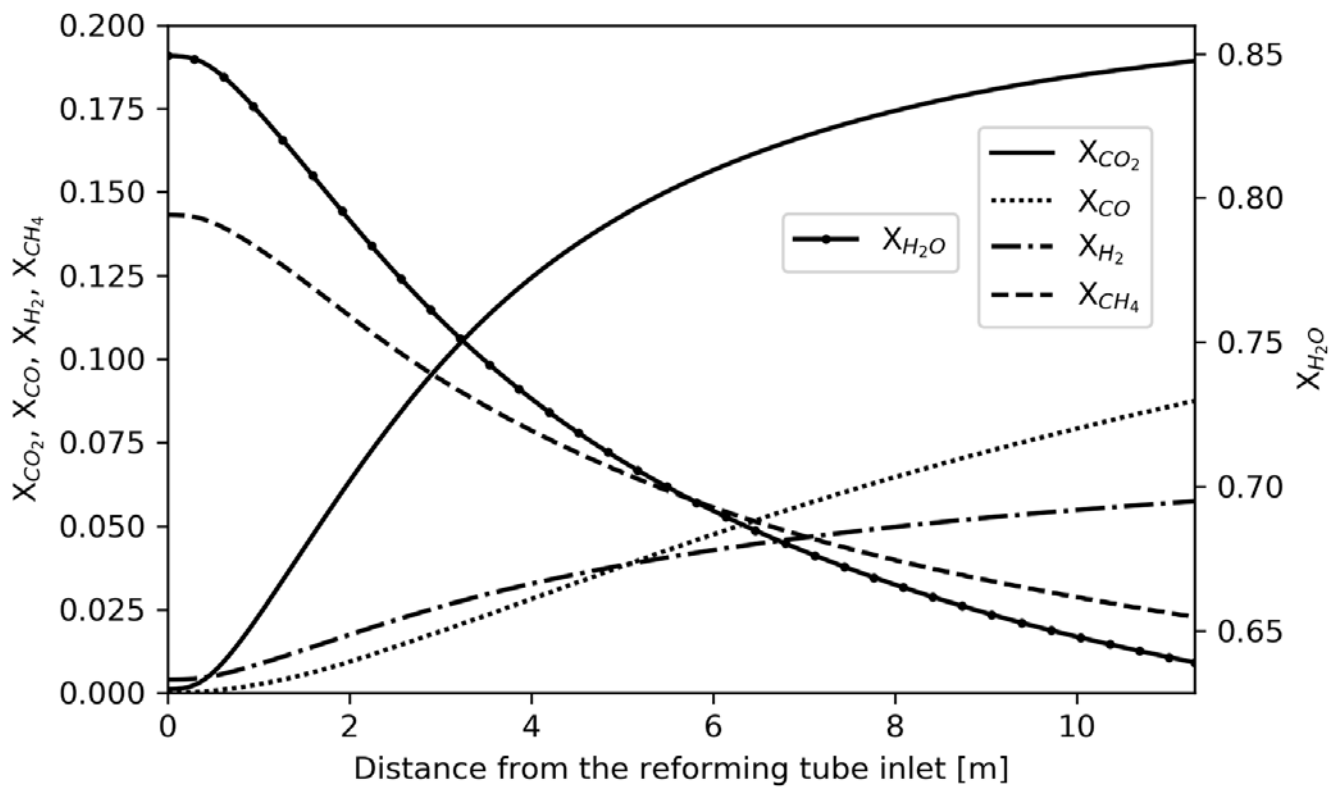

**Fig. 18.** The computed average mass-faction based composition ($X^i$) profiles in all RTs.

*4.2.4. Composition profiles*

Fig. 18 illustrates the computed composition profiles of reacting species in terms of their mass fractions for SMR process taking place inside all the RTs (results are not considered for every RT here). High gradients of compositions are clearly observed in just a few meters from the RT inlets. With a distance from the RT inlet, $CO_2$, CO, $H_2$ mass fractions gradually increase while $CH_4$ and $H_2O$ mass fractions gradually decrease. Normally, mass fractions are expected to gradually vary along the tube length rather than reaching an asymptotic value. Therefore, the present simulation case predicts this evolution characteristic well. Additionally, the numerical average methane conversion (% $CH_4$) for all RTs is calculated as 0.829 while the plant data is reported as 0.710.

## 5. Conclusions

The present study proposes an optimized modeling and simulation framework which combines fully-coupled appropriate furnace side models with process side model in a proper modeling strategy to provide a complete and accurate representation of combustion and heat transfer phenomena in the radiation section, i.e. firebox of an industrial top-fired hydrogen reforming furnace. The comparative analysis of the results obtained from different coupled modeling approaches with the experimentally measured data in terms of the average outlet temperatures of the firebox suggest that the best appropriate CFD modeling approach couples two-layer Realizable k-$\varepsilon$ turbulence model with the Flamelet Generated Manifold (FGM) model with a sub-model called Turbulent Flame Speed Closure (TFC), Discrete-Ordinates (DO) model with weighted-sum-of-gray-gases (WSGGM) radiation sub-model and a specialized 1-D plug flow Reacting Channel (RC) model. Simulation results of the selected approach are also validated with the experimentally measured data in terms of the point-wise temperature distributions at the panel walls of the furnace. The study is further extended to qualitative and quantitative analysis of the results for velocity vector field, temperature distributions at different planes of the firebox and heat flux/temperature distributions and species composition profiles in the RTs. These results demonstrate that the proposed modeling framework with the chosen coupled modeling approach can reproduce well basic flow features including combustion process, flame formation, downward movement of flue-gas in association with large recirculation zones, radiative heat transfer to the RTs and heat maps of the reforming tubes. This modeling approach hereby is proved to perform realistic computation of velocity, temperature and concentration fields in the firebox and species composition, heat flux and temperature distributions in the RTs as well as reproducing detailed heat maps, which give an insight for better analysis of furnace-balancing in terms of heat maps of the furnace. A significant heat transfer from the flue-gas to RTs within the region from the burner entrance to the high temperature region is observed and burner lane temperature fields exhibit different distribution characteristics i.e. non-uniform temperature distributions at different heights of the furnace which shows that the furnace is not balanced. In addition, results further reveal that important gradients are observed in the maximum reaction zone of the RTs i.e. along the entrance length for the mass-fraction based compositions and temperature. Mass fractions seem to gradually vary along the tube length rather than reaching an asymptotic value as expected. Therefore, the present simulation case predicts this evolution characteristic well. The results in general also suggest that the proposed coupled modeling approach can be used with confidence in accurately simulating fluid flow/heat transfer in the firebox to further optimize the geometry and operating parameters in terms of more uniform temperature distribution with higher combustion efficiency provided that there is quality experimental data for a numerical validation.

## Declaration of Competing Interest

The authors declare that they have no known competing financial interests or personal relationships that could have appeared to influence the work reported in this paper.

## CRediT authorship contribution statement

**Mustafa Tutar:** Writing – original draft. **Cihat Emre Üstün:** Software, Data curation, Writing – original draft. **Jose Miguel Campillo-Robles:** Writing – review & editing, Investigation. **Raquel Fuente:** Funding acquisition, Project administration. **Silvia Cibrián:** Validation. **Arturo Fernández:** Resources, Writing – review & editing, Supervision. **Gabriel A. López:** Writing – review & editing, Project administration.

## Acknowledgments

This research is partially funded by Basque Industry 4.0 programme of Basque Government (BI00024/2019) and University-Company-Society 2019 call of UPV/EHU (US19/13). Open access funding is provided by the University of the Basque Country (UPV/EHU).